%% file: manuscript.tex
\documentclass{article}
\usepackage[T1]{fontenc}
\usepackage{iclr2027_conference,times}
\iclrfinalcopy
\usepackage{amsmath,amssymb}
\usepackage{hyperref}
\hypersetup{hidelinks}
\usepackage{url}

\usepackage{booktabs,array,xcolor,graphicx,flafter}
\usepackage{placeins,wrapfig}
\usepackage{algorithm,algpseudocode}
\definecolor{RankRed}{HTML}{B33A3A}
\definecolor{RankBlue}{HTML}{2463A6}
\definecolor{RankGreen}{HTML}{287A52}
\newcommand{\RankFirst}[1]{{\color{RankRed}\textbf{#1}}}
\newcommand{\RankSecond}[1]{{\color{RankBlue}\underline{#1}}}
\newcommand{\RankThird}[1]{{\color{RankGreen}#1}}
\newcommand{\ResultCell}[2]{\mbox{#1{\scriptsize #2}}}
\usepackage{enumerate}
\usepackage{enumitem}
\usepackage{pifont}
\setlist[itemize]{leftmargin=*}

\title{Revisiting Temporal Link Prediction through Structural Proximity}
\input{authors}

\begin{document}
\maketitle

\begin{abstract}
\frenchspacing
Existing temporal link prediction methods typically rely on learned temporal representations to capture evolving interaction patterns. In this paper, we show that a classical structural proximity measure can match or outperform these methods without training or learned temporal sequence modeling. We revisit Personalized PageRank (PPR) on historical interaction graphs, using simple sliding-window retention or exponential time decay. Surprisingly, across 14 datasets, Time-decay PPR ranks first on nine and second on four among the compared methods, including datasets with few repeated edges. To understand this result, we examine the historical paths connecting source nodes to previously unseen destinations. We find that such paths contain useful predictive information, and that PPR outperforms a strong neural baseline on queries where historical structure clearly distinguishes the true destination from negative candidates. These results show that strong performance on current benchmarks does not necessarily require complex temporal models. More fundamentally, they raise the question of how much predictive value learned temporal modeling adds beyond historical structural proximity. Our findings motivate reassessing progress in temporal link prediction with strong classical baselines.
\end{abstract}

\section{Introduction}
Temporal graphs represent relationships between entities through interactions over time and are widely used in social networks, recommender systems, and financial transactions. A central challenge is temporal link prediction, which aims to predict which node pairs are likely to interact in the future based on observed interactions. Accurate predictions support applications such as item recommendation and social recommendations as users' interests and behaviors evolve. Future interactions may either recur between previously connected node pairs or establish new relations, requiring models to exploit historical information while generalizing beyond direct interaction history.

Existing temporal graph learning methods learn representations through dynamic node states, temporal neighborhood aggregation, or historical sequence encoding \citep{rossi2020tgn,xu2020tgat,cong2023graphmixer,yu2023dygformer}. In contrast, EdgeBank shows that simply memorizing historical edges can already achieve strong performance when interactions frequently recur, particularly under evaluation settings with easily distinguishable negative candidates~\citep{poursafaei2022evaluation}. 
This observation has motivated benchmarks that place greater emphasis on prediction beyond direct interaction memorization. TGB broadens evaluation across diverse domains and more realistic protocols~\citep{huang2023tgb}, while TGB-Seq introduces datasets with fewer repeated edges to challenge models to capture sequential dynamics and generalize to unseen relations \citep{yi2025tgbseq}. 
More recently, CRAFT matches candidate destinations against the source's recent interaction history using learnable node embeddings and target-aware cross-attention, achieving strong performance with a comparatively simple architecture~\citep{yi2025craft}. 

\begin{figure}[t]
\centering
\setlength{\abovecaptionskip}{-3mm}
\setlength{\belowcaptionskip}{-1mm}
\includegraphics[width=\linewidth]{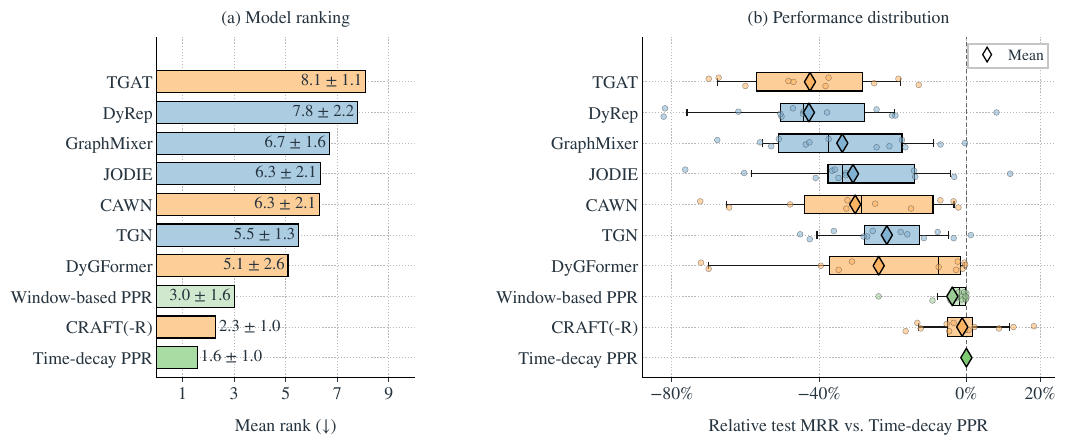}
\caption{(a) Mean rank over each method's reported results across 14 datasets; missing results are excluded. Standard deviations are across datasets, and ties share the same rank. (b) Relative test MRR against Time-decay PPR across available results from 14 datasets. Dots represent dataset results, boxes indicate interquartile ranges, whiskers mark the 10th--90th percentiles, and diamonds denote means. Negative values favor Time-decay PPR.}
\label{fig:main-comparison}
\end{figure}

However, reducing repeated edges does not remove the structural information contained in the historical graph. A source and a previously unseen destination may never have interacted directly while still being connected through historical multi-hop paths. This structural view has a long history in classical link prediction. 
Common Neighbors and Adamic--Adar score node pairs through local neighborhood overlap, while Katz and rooted PageRank capture multi-step connections \citep{libennowell2007linkprediction}. 
Personalized PageRank (PPR) provides a natural probe of structural proximity: starting from a query source, it aggregates random walks of different lengths and therefore assigns scores not only to previous neighbors but also to nodes reachable through multi-hop paths \citep{gleich2015pagerank}. Consequently, a destination can receive a high PPR score even when it has never directly interacted with the source. This raises a more basic question: \textbf{how much predictive information does historical multi-hop structure provide for future link prediction?}

To investigate this question, we revisit PPR as a controlled, training-free baseline for measuring the predictive value of historical structural proximity. To accommodate evolving interaction streams, we construct historical graphs using simple temporal retention mechanisms. Window-based PPR keeps only recent interactions, while Time-decay PPR gradually downweights older events. As shown in Figure~\ref{fig:main-comparison}, Window-based PPR performs competitively with existing temporal graph models, showing that multi-hop structure in a recent historical graph provides a strong predictive signal. 
More surprisingly, Time-decay PPR ranks first on nine of the 14 datasets and second on another four among the compared methods, including datasets with few repeated edges. 

Further analysis connects this performance to how historical structure evolves. New interactions often close existing multi-hop paths into cycles, suggesting that future links are frequently foreshadowed by structural connectivity already present in the historical graph. At the query level, historical path patterns can distinguish the true destination from negative candidates, and PPR outperforms a strong neural baseline on queries where historical structural signals are particularly discriminative. These findings show that historical structural proximity remains highly predictive even when direct edge recurrence is limited. This does not imply that temporal modeling is unnecessary. Instead, our results suggest that future progress should be evaluated relative to the structural signal already available in historical graphs. This motivates evaluation protocols that better control for structural proximity and targeted analysis of queries where historical paths are weak or ambiguous. They also suggest developing models that focus temporal learning on information not already explained by historical connectivity. 
Our main contributions are as follows:
\begin{itemize}
    \item We systematically revisit PPR as a training-free structural baseline for future link prediction. Across 14 datasets, Time-decay PPR ranks first on nine and second on four, while PPR without temporal decay also remains competitive.
    \item We analyze the predictive value of historical structural proximity from the perspective of graph evolution. By examining cycle closure, historical path discriminability, and model predictions, we study how indirect historical connections support unseen-link prediction and how effectively models use this evidence.
    \item We disentangle the effects of historical retention, interaction recency, and repeated interactions, and use these analyses to reassess how much predictive value learned temporal modeling adds beyond historical graph structure under current evaluation protocols.
\end{itemize}

\section{Preliminaries and Related Work}
\subsection{Problem Formulation}
\label{sec:task-formulation}
\paragraph{Temporal Graph.}
A temporal graph \(\mathcal G=(\mathcal V,\mathcal E)\) consists of a node set \(\mathcal V\) and a sequence of timestamped interactions \(\mathcal E\), ordered by non-decreasing time. Each interaction \(e_i=(u_i,v_i,t_i)\) records an event between source node \(u_i\) and destination node \(v_i\) at timestamp \(t_i\), such as a purchase or a message. 
Multiple interactions may occur between the same source and destination at different timestamps.

\paragraph{Future Link Prediction.}
Given a query source \(s\), query time \(t\), and the interaction history \(\mathcal H_t=\{(u_i,v_i,t_i)\in\mathcal E\mid t_i<t\}\), future link prediction aims to identify the destination \(d\) that interacts with \(s\) at time \(t\). Following the evaluation protocols of TGB and TGB-Seq, we formulate this task as candidate ranking \citep{huang2023tgb,yi2025tgbseq}. A candidate destination set \(\mathcal C\) contains the true destination \(d\) and a set of negative candidates. Using only \(\mathcal H_t\), a predictor assigns a score \(f(s,v,t)\) to each candidate \(v\in\mathcal C\), with the objective of ranking the true destination ahead of the negative candidates. 

\subsection{Related Work}
\label{sec:related-work}
Existing approaches to future link prediction exploit historical interactions through both structural proximity and learned temporal representations.

\paragraph{Structural Proximity and Personalized PageRank.} 
Structural proximity methods score node pairs directly from neighborhood overlap and multi-step connections in historical networks. Common Neighbors and Adamic--Adar capture local associations through shared neighbors, whereas Katz and rooted PageRank account for connections or walks of different lengths. \citet{libennowell2007linkprediction} systematically compared these measures, demonstrating the predictive value of historical topology for new links. PPR produces rankings personalized to a node or a set of nodes and has also been used for personalized search and local community detection \citep{jeh2003scaling,avron2015community}.

Subsequent work has explored incremental ranking updates as graphs change and their use in tracking local communities \citep{zhang2016approximate,nathan2017local}, as well as incorporating diffusion into graph representation learning. PPNP/APPNP~\citep{gasteiger2019appnp} combines node prediction with PPR propagation, while GPR-GNN~\citep{chien2021gprgnn} learns weights for different propagation orders. PPR thus serves both as a direct measure of structural proximity and as a propagation component in learned models.

\paragraph{Dynamic Graph Representation Learning.} 
Dynamic graph representation learning extracts predictive information through node-state updates, temporal neighborhood aggregation, and historical sequence encoding. JODIE, DyRep, and TGN maintain node states that evolve with interaction events \citep{kumar2019jodie,trivedi2019dyrep,rossi2020tgn}. TGAT, GraphMixer, and DyGFormer learn node associations from time-aware neighborhoods or interaction sequences \citep{xu2020tgat,cong2023graphmixer,yu2023dygformer}. CRAFT instead predicts interactions by matching candidate destinations against the source's recent history \citep{yi2025craft}.

Beyond learned representations, EdgeBank~\citep{poursafaei2022evaluation} shows that a training-free memory of historical edges can perform strongly when interactions frequently recur, exposing limitations of evaluation settings with easily distinguishable negative candidates. PopTrack tracks recent destination popularity and highlights the influence of negative sampling on temporal graph evaluation \citep{daniluk2023temporal}. Recency- and frequency-based heuristics can also match or outperform neural models on several temporal graph benchmarks \citep{cornell2025heuristics}. TGB~\citep{huang2023tgb} extends evaluation to large-scale datasets from diverse domains and more challenging candidate-ranking protocols, while TGB-Seq~\citep{yi2025tgbseq} further emphasizes datasets with few repeated edges, highlighting the difficulty of predicting unseen relations and capturing sequential dynamics. These developments motivate revisiting a complementary question: how much predictive information remains in the historical graph structure itself under modern future link prediction protocols?

\section{Revisiting PPR for Temporal Link Prediction}
\label{sec:ppr}

In this section, we describe how we use PPR \citep{brin1998anatomy,gleich2015pagerank} to measure the predictive value of historical graph structure for future link prediction. We first formulate finite-step PPR as a multi-hop structural proximity measure~\citep{gasteiger2019appnp}, then describe how timestamped interactions are converted into historical graphs through temporal retention and repeated-interaction aggregation. Finally, we use the resulting PPR scores to rank candidate destinations.

\subsection{PPR as Multi-hop Structural Proximity}
\label{sec:finite-step-ppr}
Personalized PageRank measures proximity to a source node through a random walk with restart. For a transition matrix $\mathbf{P}$ and source one-hot row vector $\mathbf{e}_s$, the stationary PPR vector $\boldsymbol{\pi}_s$ satisfies
\begin{equation}
\boldsymbol{\pi}_s
=
\alpha \mathbf{e}_s
+
(1-\alpha)\boldsymbol{\pi}_s\mathbf{P},
\end{equation}
where $\alpha\in(0,1)$ is the restart probability. At each step, the walk returns to the source with probability $\alpha$ and follows an outgoing graph transition with probability $1-\alpha$. We use a finite-step approximation initialized by $\mathbf{p}_s^{(0)}=\mathbf{e}_s$:
\begin{equation}
\mathbf{p}_s^{(\ell+1)}
=
\alpha\mathbf{e}_s
+
(1-\alpha)\mathbf{p}_s^{(\ell)}\mathbf{P},
\qquad
\ell=0,\ldots,L-1.
\end{equation}
After $L$ iterations,
\begin{equation}
\mathbf{p}_s^{(L)}
=
\alpha\sum_{\ell=0}^{L-1}
(1-\alpha)^\ell
\mathbf{e}_s\mathbf{P}^{\ell}
+
(1-\alpha)^L
\mathbf{e}_s\mathbf{P}^{L}.
\end{equation}
Each term $\mathbf{e}_s\mathbf{P}^{\ell}$ captures connectivity through walks of length $\ell$. Therefore, finite-step PPR combines direct and multi-hop structural connections up to $L$ propagation steps, while $\alpha$ controls their relative contributions. Importantly, a candidate that has never directly interacted with the source can still receive a nonzero score when it is connected to the source through historical multi-hop paths. We also include $\alpha=0$, for which the finite-step iteration reduces to an $L$-step random walk without restart, with scores given by $\mathbf{e}_s\mathbf{P}^{L}$.

\subsection{Historical Graph Construction}
\label{sec:historical-graph}

Applying PPR to temporal link prediction requires representing interactions of different ages and repeated occurrences in a historical graph. We construct an undirected weighted graph from interactions observed before the query time $t$, without adding self-loops. Temporal retention determines how each event is weighted, and repeated-interaction aggregation combines events between the same node pair. Let $\mathcal{T}_{uv}(t)$ denote the multiset of interaction timestamps between $u$ and $v$ before $t$.

\paragraph{Temporal retention.}
We assign each interaction a weight based on its age $\Delta t$, using either exponential decay with half-life $h>0$ \citep{zuo2018htne,qiu2020hnip,lu2024tpnet} or a sliding window of length $W>0$ \citep{poursafaei2022evaluation}:
\begin{equation}
\kappa_h^{\mathrm{decay}}(\Delta t)=2^{-\Delta t/h},
\qquad
\kappa_W^{\mathrm{window}}(\Delta t)=\mathbb{I}[\Delta t\leq W].
\end{equation}
Exponential decay emphasizes recent interactions while preserving older connections. Sliding windows impose a hard cutoff that can remove connections supporting multi-step propagation. We refer to the corresponding PPR variants as Time-decay PPR and Window-based PPR, respectively. Both include full-history retention as $h$ or $W$ tends to infinity.

\paragraph{Repeated-interaction aggregation.}
For either retention function $\kappa$, we combine events from the same node pair as
\begin{equation}
w_{uv}(t)=\mathcal{A}\bigl(\{\kappa(t-\tau):\tau\in\mathcal{T}_{uv}(t)\}\bigr),
\qquad \mathcal{A}\in\{\max,\mathrm{sum}\},
\end{equation}
where the braces denote a multiset. Max aggregation retains the latest interaction's weight, yielding last-decay or binary-window weighting; sum aggregation accumulates all event weights, yielding sum-decay or count-window weighting. These choices distinguish the recency or presence of a relation from the accumulated evidence of its recurrence. The weights define a symmetric adjacency matrix $\mathbf{A}_t$, which we row-normalize to obtain $\mathbf{P}_t$.

\subsection{Finite-Step Diffusion and Candidate Scoring}
\label{sec:candidate-scoring}

Given the historical transition matrix $\mathbf{P}_t$, we run finite-step PPR from the query source $s$. For each candidate destination $v\in\mathcal{C}$, we use $f(s,v,t)=[\mathbf{p}_t^{(L)}(s)]_v$ as its prediction score and rank candidates in descending order, following source-rooted PageRank for link prediction \citep{libennowell2007linkprediction}.

\section{Experiments}
\label{sec:experiments}

\subsection{Experimental Setup}
\label{sec:experimental-setup}

\textbf{Datasets and baselines.} We evaluate 14 datasets from TGB, TGB-Seq, and established temporal graph benchmarks, grouped into seen-dominant and unseen-dominant regimes following \citet{yi2025craft}. Their domains, sizes, and repeated-edge ratios are summarized in Appendix~\ref{app:dataset-statistics}. We compare with the reported results of JODIE, DyRep, TGAT, TGN, CAWN, GraphMixer, DyGFormer, and CRAFT, using CRAFT-R for the seen-dominant group and CRAFT for the unseen-dominant group \citep{yi2025craft}. Model references and configuration details are in Appendix~\ref{app:experimental-details}. We additionally evaluate EdgeBank \citep{poursafaei2022evaluation} and heuristics based on pairwise interaction history and destination recency, frequency, and activity; definitions and results are provided in Appendix~\ref{app:historical-baselines}.

\textbf{Evaluation and selection.} We use chronological training, validation, and test splits and evaluate candidate ranking by Mean Reciprocal Rank (MRR). Released split assignments and benchmark-provided negative candidates are retained where available \citep{huang2023tgb,yi2025tgbseq}; other candidate sets follow their corresponding ranking protocols. For both Time-decay PPR and Window-based PPR, we select configurations by validation MRR before test evaluation; the search space is detailed in Appendix~\ref{app:experimental-details}. PPR scores each timestamp group or timestamp-aligned batch before updating the historical graph with its positive interactions; refresh settings are detailed in Appendix~\ref{app:experimental-details}.

\subsection{Main Results}
\label{sec:main-results}

Time-decay PPR achieves the best test MRR on 9 of the 14 datasets and ranks second on another 4, including ties (Tables~\ref{tab:main-unseen} and~\ref{tab:main-seen}). Without learning node representations or prediction parameters, it remains competitive on both seen-dominant and unseen-dominant datasets. Window-based PPR is also competitive, but Time-decay PPR matches or exceeds it on every dataset. We examine the role of temporal retention in Section~\ref{sec:temporal-retention-analysis}.

\input{tables/main_results}

On seen-dominant datasets where Time-decay PPR outperforms the strongest reported baselines, its relative MRR gains are generally modest, reaching approximately 2.9\% on UCI and 2.4\% on Flights. More strikingly, it remains highly competitive on unseen-dominant datasets, where direct edge recurrence is less prevalent. Its relative MRR gains over the strongest reported baselines reach approximately 19.8\% on YouTube, 15.4\% on Flickr, and 14.2\% on WikiLink, with smaller gains of 5.7\% on GoogleLocal and 1.6\% on Taobao. These results motivate our analysis in Section~\ref{sec:analysis} of how historical paths and structural proximity support the prediction of unseen relations.

Additional comparisons in Appendix~\ref{app:historical-baselines} show that pair-recency and pair-frequency perform strongly on several seen-dominant datasets. By contrast, destination activity outperforms these pair-based heuristics on GoogleLocal, YouTube, and Flickr and approaches the strongest reported neural baseline on tgbl-review. These results highlight the dataset-dependent effectiveness of simple historical statistics. Time-decay PPR, however, remains competitive across both seen- and unseen-dominant regimes, suggesting that historical structural proximity provides useful predictive information across diverse interaction patterns. We further evaluate activity-weighted negative sampling on five datasets. Time-decay PPR remains competitive under activity-weighted candidate sets; complete results and analysis are provided in Appendix~\ref{app:activity-weighted-negative-sampling}.

\section{Analysis}
\label{sec:analysis}

In this section, we analyze how historical structural proximity supports temporal link prediction. We first examine cycle formation in temporal graphs relative to a random reference, then study how often new links close historical paths and whether these paths distinguish positive from negative candidates. We further investigate how effectively a strong neural baseline uses this structural evidence. Finally, we examine the effects of temporal retention, repeated-interaction aggregation, and propagation.

\subsection{Structural Evolution and Its Predictive Value}
\label{sec:structural-discriminability}

\begin{figure}[!htb]
\centering
\includegraphics[width=0.9\linewidth,trim=35bp 110bp 35bp 70bp,clip]{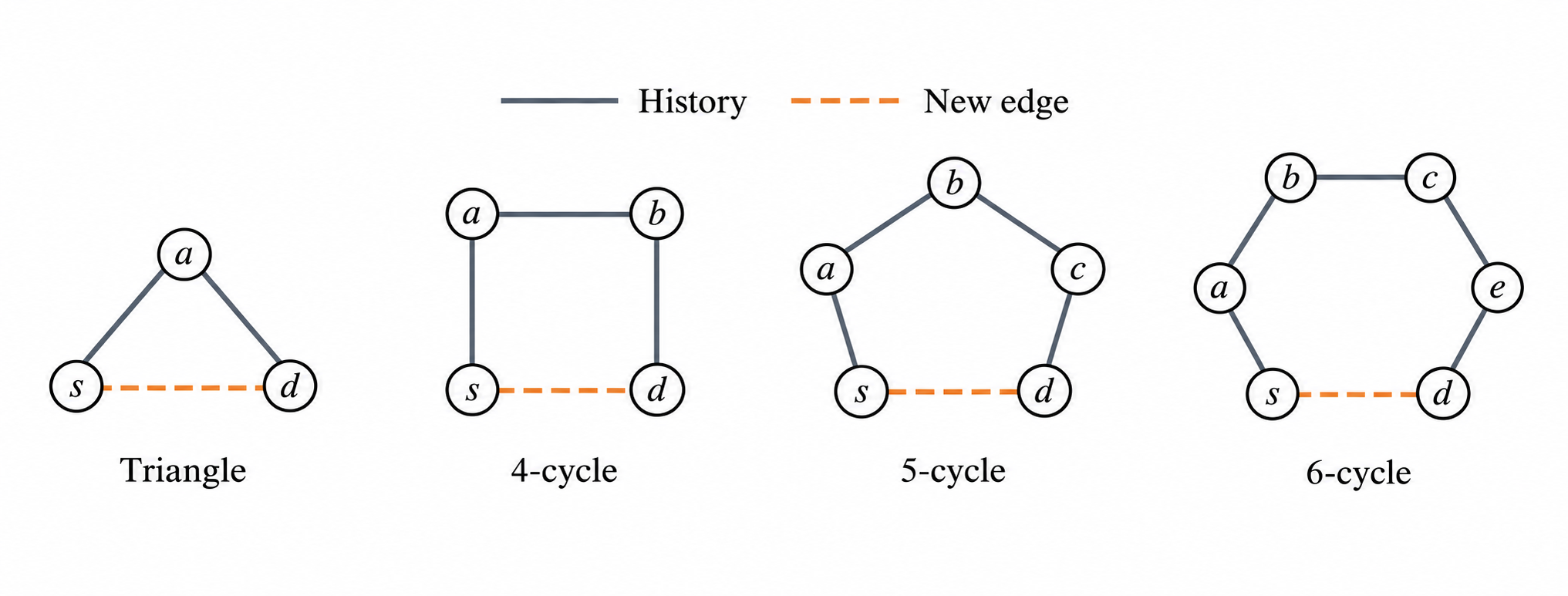}
\caption{A new edge closes a historical path into a cycle.}
\label{fig:cycle-closure-schematic}
\end{figure}

PPR's strong performance on unseen-dominant datasets motivates examining how historical structural proximity supports the prediction of unseen relations. New interactions may connect nodes already linked by historical paths, closing these paths into cycles (Figure~\ref{fig:cycle-closure-schematic}). We first ask whether such cycles accumulate earlier in real interaction sequences than under random edge ordering.

Motivated by randomized-reference models for temporal networks \citep{gauvin2022randomized}, we use the following reference.

\noindent\textbf{Definition (Random-order reference).} Let $G_M=(V,E_M)$ be the final aggregated simple graph with $M$ distinct edges. Draw a uniformly random permutation of $E_M$, and let $\widetilde G_m$ contain its first $m$ edges. The expected number of simple $k$-cycles is
\begin{equation}
\mathbb{E}[N_k(\widetilde G_m)\mid G_M]
=N_k(G_M)\frac{\binom{m}{k}}{\binom{M}{k}}.
\label{eq:random-order-cycle-expectation}
\end{equation}
Across seven datasets, observed four- and six-cycle counts exceed this expectation at every measured intermediate snapshot, indicating earlier cycle accumulation under the real ordering. Appendix~\ref{app:cycle-growth} provides the derivation and comparison curves.

\begin{figure}[!htb]
\centering
\includegraphics[width=\linewidth]{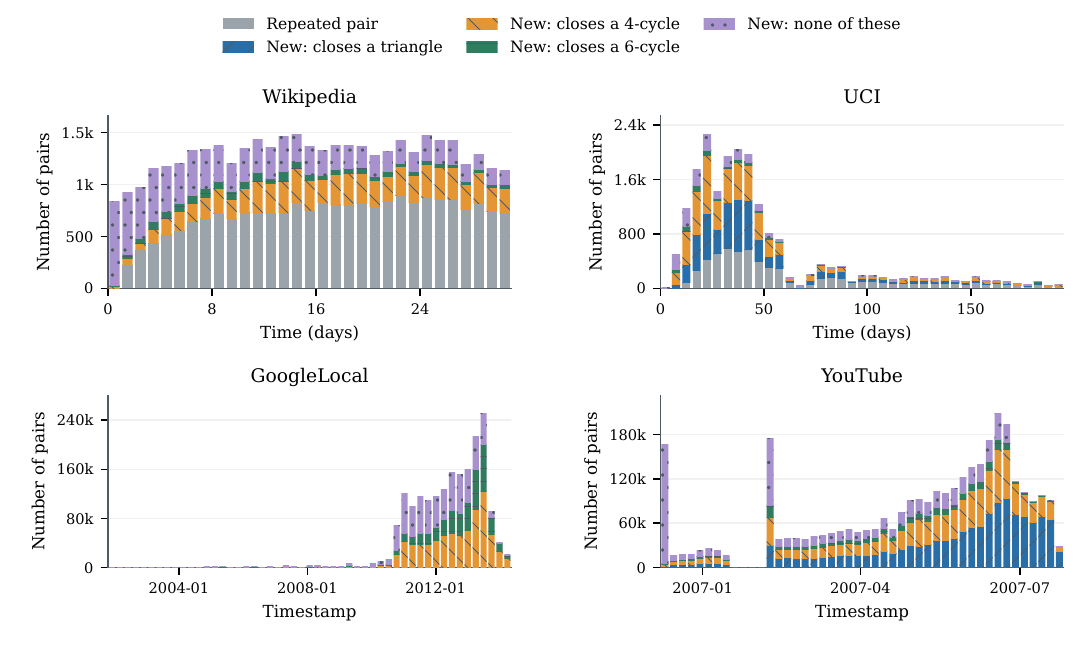}
\caption{Unique node pairs per time interval, grouped by recurrence and closure of historical triangles, four-cycles, or six-cycles. Earlier categories take precedence; definitions are in Appendix~\ref{app:structural-settings}.}
\label{fig:event-structural-appearance}
\end{figure}

To connect this evolution pattern to incoming interactions, each new edge closes a $k$-cycle for every existing $(k-1)$-step simple path between its endpoints. Inspired by EdgeBank's temporal edge appearance plots \citep{poursafaei2022evaluation}, Figure~\ref{fig:event-structural-appearance} extends the repeated/new distinction by grouping new pairs according to cycle closure. Each node pair is counted once per display interval. Wikipedia exhibits substantial recurrence, while new relations on UCI, GoogleLocal, and YouTube often close historical paths. Paths are defined on the aggregated historical graph without requiring chronological order among their interactions.

Historical paths provide predictive evidence when they distinguish positive from negative candidates. Figure~\ref{fig:structural-coverage-all} in Appendix~\ref{app:structural-settings} compares path coverage on complete test sets, using prediction-time graphs and original candidates. Wikipedia and UCI show strong one-step coverage gaps; GoogleLocal and YouTube have near-zero one-step coverage but clear three-step and two-step gaps, respectively. Table~\ref{tab:path-discriminability-scores} further evaluates ranking by unweighted simple-path counts. For previously unconnected nodes in a graph without self-loops, every three-step walk is a simple path, linking this diagnostic to PPR's transition-weighted three-step component. We next examine how effectively PPR and a strong neural baseline use this evidence.

\subsection{Why Do Neural Models Fall Short?}
\label{sec:neural-structural-gap}

We compare Time-decay PPR with CRAFT \citep{yi2025craft}, a strong neural baseline using target-aware matching of recent interaction histories, on GoogleLocal and YouTube. Test queries are grouped by how well historical paths distinguish positive from negative candidates. In the strong-structure group, the true destination has strictly more three-step simple paths from the source than every negative candidate. Path count alone therefore ranks it uniquely first. The remaining queries form medium-structure, weak-structure, and no-three-step-support groups, defined in Appendix~\ref{app:structural-group-decomposition}. Grouping uses historical path counts rather than model scores.

\begin{figure}[!htb]
\begin{minipage}[t]{0.48\linewidth}
\vspace{0pt}
\centering
\setlength{\parskip}{0pt}
\setlength{\abovecaptionskip}{0pt}
\setlength{\belowcaptionskip}{4pt}
\caption{Test MRR by structural group; percentages denote query shares.}
\label{fig:strong-structure-comparison}
\includegraphics[width=\linewidth]{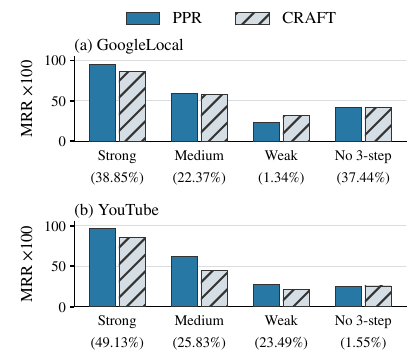}
\end{minipage}\hfill
\begin{minipage}[t]{0.48\linewidth}
\vspace{0pt}
\setlength{\parskip}{0pt}
\setlength{\abovecaptionskip}{0pt}
\setlength{\belowcaptionskip}{4pt}
\makeatletter\def\@captype{table}\makeatother
\renewcommand{\arraystretch}{1.35}
\input{tables/retention_four_arms}
\end{minipage}
\end{figure}

Within the strong-structure group, PPR achieves relative MRR improvements of approximately \textbf{9.3\% and 13.5\%} over the evaluated CRAFT system on GoogleLocal and YouTube, respectively (Figure~\ref{fig:strong-structure-comparison}). These queries account for \textbf{38.85\% and 49.13\%} of the respective test sets. When weighted by their query proportions, the within-group performance gaps account for nearly all of PPR's overall net MRR gain on GoogleLocal and approximately half on YouTube (Table~\ref{tab:strong-structure-decomposition}). These results suggest that modeling historical interaction sequences does not necessarily ensure effective use of higher-order structural information for link prediction.

\subsection{Temporal Retention, Aggregation, and Propagation}
\label{sec:temporal-retention-analysis}

We next examine how historical graph construction and propagation parameters shape PPR's use of the structural evidence identified above.

\begin{figure}[!htb]
\centering
\setlength{\parskip}{0pt}
\setlength{\abovecaptionskip}{4pt}
\includegraphics[width=0.90\linewidth]{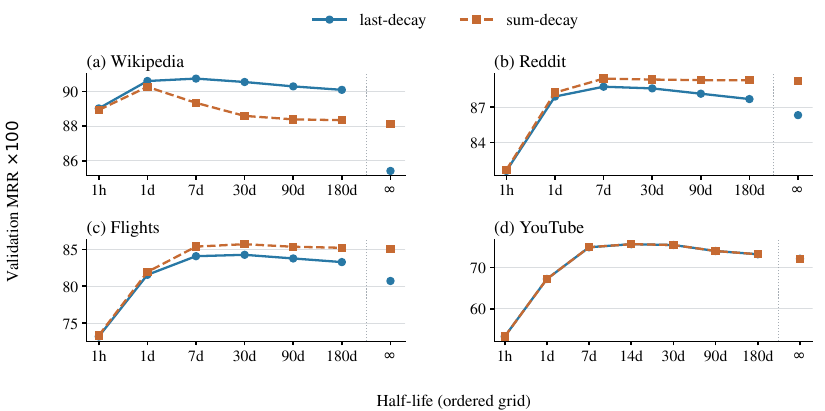}
\caption{Validation sensitivity to half-life and aggregation, with $\alpha$ and propagation depth fixed. Half-lives follow an ordered grid; $\infty$ denotes no decay. Each panel has its own vertical scale.}
\label{fig:section63-temporal}
\end{figure}

\textbf{Temporal retention and repeated interactions.} Table~\ref{tab:retention-four-arms} compares exponential decay, sliding-window retention, and full-history retention without decay. For the window ablation, we retain the aggregation and propagation parameters of Time-decay PPR and select only the window length using validation MRR. Exponential decay achieves the best or tied-best performance across all eight datasets among the evaluated configurations. This suggests that temporal link prediction benefits from emphasizing recent interactions while retaining temporally weighted long-term history. Figure~\ref{fig:section63-temporal} further shows that an appropriate half-life outperforms both an overly short half-life and no decay. The preferred time scale varies across datasets, motivating adaptive time scales and multi-scale history modeling.

Figure~\ref{fig:section63-temporal} also illustrates the joint effects of half-life and repeated-interaction aggregation. On unseen-dominant datasets such as YouTube and Flickr, the low prevalence of repeated interactions results in little difference between last-decay and sum-decay. On seen-dominant datasets, however, accumulating repeated interactions offers no consistent advantage: across the evaluated finite half-lives, last-decay consistently performs better on Wikipedia, whereas sum-decay consistently performs better on Reddit and Flights. These findings suggest that the choice of aggregation should account for the interaction patterns and temporal distribution of repeated events. Complete curves for seven datasets, including Flickr, are provided in Appendix~\ref{app:section63-settings}.

\begin{figure}[!htb]
\centering
\setlength{\parskip}{0pt}
\setlength{\abovecaptionskip}{4pt}
\includegraphics[width=0.90\linewidth,trim=0bp 11bp 0bp 4bp,clip]{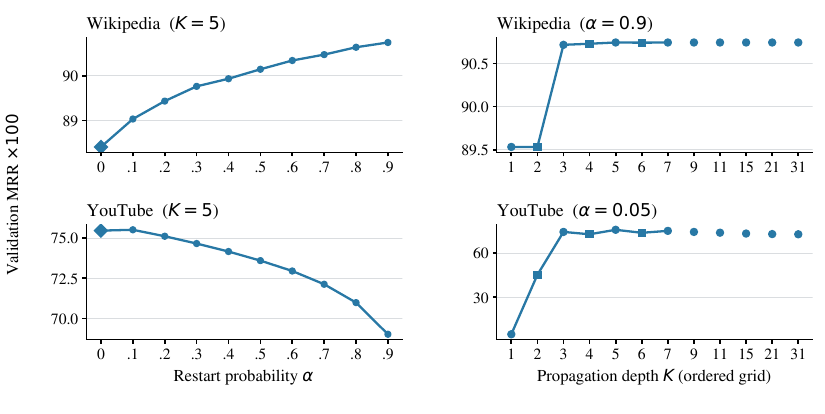}
\caption{Validation sensitivity to $\alpha$ and $K$ on Wikipedia and YouTube. Other parameters are fixed. $K$ uses an ordered grid; markers show measured values. Each panel has its own vertical scale.}
\label{fig:section63-propagation}
\end{figure}

\textbf{Propagation parameters.} Figure~\ref{fig:section63-propagation} shows how propagation depth $K$ (the iteration count $L$ in Section~\ref{sec:finite-step-ppr}) and restart probability $\alpha$ affect prediction performance. On most datasets, shallow propagation achieves or approaches the best performance. One-step propagation already performs well on seen-dominant datasets, whereas unseen-dominant datasets typically obtain their main gains at $K=3$ to $5$; increasing the depth further does not consistently improve performance. Preferred values of $\alpha$ also vary across datasets, motivating further exploration of learnable weights over propagation orders. All sensitivity results use validation data; additional datasets and fixed settings are reported in Appendix~\ref{app:section63-settings}.

\FloatBarrier
\section{Conclusion}
In this paper, we revisit temporal link prediction through structural proximity and reassess classical Personalized PageRank as a training-free baseline. With appropriate temporal retention and repeated-interaction aggregation, the resulting Time-decay PPR achieves strong performance across widely used benchmarks. Our analyses connect this performance to historical structural proximity, showing that multi-step historical paths provide useful predictive evidence even for previously unseen relations. Comparisons with a strong neural baseline further reveal a gap between the structural evidence available and its use in prediction. These findings motivate temporal graph models that better exploit historical structural proximity and highlight the importance of strong classical baselines when evaluating progress in temporal link prediction.
\FloatBarrier

\subsection*{AI use statement}

In this work, we used generative AI tools to assist with writing and language polishing, as well as the implementation of part of the code. We did not use generative AI tools to design experiments, assist with proofs, interpret results, or propose research hypotheses. The LLM-generated code was verified and tested for correctness by two authors. We take responsibility for the final content of this work, including text, claims, and artifacts produced with the aid of generative AI.
\section*{Reproducibility Statement}
Section~\ref{sec:ppr} describes the PPR variants, and Section~\ref{sec:experimental-setup} specifies the evaluation protocol. Appendix~\ref{app:experimental-details} provides hyperparameter settings and graph-update procedures. Additional baselines and candidate construction are detailed in Appendix~\ref{app:additional-baselines-candidates}.

\bibliographystyle{iclr2027_conference}
\bibliography{references}

\clearpage
\appendix
\makeatletter
\setlength{\@fptop}{0pt}
\makeatother
\section{Dataset Statistics}
\label{app:dataset-statistics}
\input{tables/dataset_statistics}

Wikipedia, Reddit, and LastFM record page edits, subreddit interactions, and music listening, respectively \citep{kumar2019jodie}. UCI contains user messages \citep{panzarasa2009patterns}, and Flights records flights between airports \citep{poursafaei2022evaluation}. TGB-Seq provides the benchmark versions of GoogleLocal, Yelp, and Taobao (business and product interactions), YouTube and Flickr (social links), and WikiLink (hyperlinks) \citep{yi2025tgbseq}. Their underlying sources include Google Local Reviews \citep{li2022uctopic,yan2023personalized}, the Yelp Open Dataset, and the Taobao User Behavior dataset \citep{zhuo2020tree}; Flickr \citep{cha2009flickr}, YouTube \citep{mislove2007online}, and WikiLink \citep{boldi2004ubicrawler,boldi2011layered} are traced to the cited network collections. The tgbl-review, tgbl-coin, and tgbl-flight datasets are from TGB and cover product reviews, financial transactions, and flights, respectively \citep{huang2023tgb}. Table~\ref{tab:dataset-statistics} reports dataset sizes and recurrence rates as reported by \citet{yi2025craft}.

\section{Experimental Details}
\label{app:experimental-details}

The main tables use baseline results reported by \citet{yi2025craft}. The compared methods are JODIE \citep{kumar2019jodie}, DyRep \citep{trivedi2019dyrep}, TGAT \citep{xu2020tgat}, TGN \citep{rossi2020tgn}, CAWN \citep{wang2021caw}, GraphMixer \citep{cong2023graphmixer}, DyGFormer \citep{yu2023dygformer}, and CRAFT \citep{yi2025craft}. These cover recurrent node states, temporal neighborhood aggregation, temporal walks, and sequence modeling.

For TGB-Seq, we retain the released split assignments. Final event proportions can differ from 70\%/15\%/15\% after preprocessing. All graph and propagation parameters are selected on validation data before test evaluation.

\subsection{Hyperparameter Selection}

For the neural baselines evaluated in our experiments, we use the best hyperparameter configurations provided by TGB-Seq and CRAFT \citep{yi2025tgbseq,yi2025craft}.

We select the configuration of each PPR variant independently using validation MRR. Time-decay PPR uses last or sum aggregation, while Window-based PPR uses binary or count aggregation. Both share the following propagation and time-scale grids:
\[
\begin{aligned}
\alpha &\in \{0,0.05,0.1,0.2,0.3,0.4,0.5,0.6,0.7,0.8,0.9\},\\
K &\in \{1,2,3,4,5,6,7,9,11,15,21,31\},\\
h,W &\in \{1\,\mathrm{h},1\,\mathrm{d},7\,\mathrm{d},14\,\mathrm{d},30\,\mathrm{d},90\,\mathrm{d},180\,\mathrm{d},\infty\}.
\end{aligned}
\]
Here, $K$ is the iteration count $L$ in Section~\ref{sec:finite-step-ppr}, and $\infty$ denotes full-history retention. Without decay, last and sum reduce to binary and count weights, respectively. Table~\ref{tab:selected-hyperparameters} lists the configurations used in the main tables. Unlike the independently selected main-table configurations, the window ablation in Section~\ref{sec:temporal-retention-analysis} varies the window length while retaining the aggregation and propagation settings of Time-decay PPR.

\input{tables/selected_hyperparameters}

\paragraph{Historical-state refresh.}
We use event batches of 2,048 interactions on Taobao, Yelp, and tgbl-coin, and 256 on GoogleLocal; the remaining datasets refresh after each timestamp group. Batches retain complete timestamp groups, and all queries in a batch are scored from its starting history before its positive interactions are added.

\begin{algorithm}[!htb]
\caption{Incremental historical graph maintenance for Time-decay PPR}
\label{alg:incremental-ppr}
\small
\begin{algorithmic}[1]
\Require Initial history $\mathcal H_0$; chronological query batches $\{B_j\}$ retaining complete timestamp groups; aggregation $g\in\{\mathrm{last},\mathrm{sum}\}$; $h,\alpha,K$
\State Set $\lambda=\ln 2/h$ (zero if $h=\infty$) and fix reference time $t_0$
\State Initialize pair log-weights from $\mathcal H_0$ using lines~\ref{alg:last-update}--\ref{alg:sum-update}
\State Build the symmetric sparse graph and row-normalized transition matrix $P$
\For{each batch $B_j$}
    \For{each query $(s,t,\mathcal C)$ in $B_j$}
        \State Score $\mathcal C$ by $K$-step PPR from $s$ using the fixed matrix $P$
    \EndFor
    \State $R\gets\varnothing$; $\mathrm{newPair}\gets\mathrm{false}$
    \For{each observed positive interaction $(u,v,\tau)$ in $B_j$ with $u\ne v$}
        \State Treat $\{u,v\}$ as an unordered pair
        \State $\mathrm{newPair}\gets\mathrm{newPair}\lor[\{u,v\}\text{ has not appeared}]$
        \State $a_{uv}\gets\max\{a_{uv},\lambda(\tau-t_0)\}$ \label{alg:last-update}
        \State $b_{uv}\gets\operatorname{logaddexp}(b_{uv},\lambda(\tau-t_0))$ \label{alg:sum-update}
        \State $R\gets R\cup\{u,v\}$
    \EndFor
    \State Use $a_{uv}$ if $g=\mathrm{last}$, otherwise $b_{uv}$, as each pair's log-weight
    \If{$\mathrm{newPair}$}
        \State Rebuild the sparse topology and row-normalize all rows of $P$
    \Else
        \State Renormalize all incident edge weights in each row $u\in R$ of $P$
    \EndIf
\EndFor
\end{algorithmic}
\end{algorithm}

\paragraph{Incremental graph maintenance.}
Our incremental implementation maintains each node pair's event count, latest timestamp, and accumulated exponential weight. For half-life $h$, let $\lambda=\ln 2/h$ and choose a fixed reference time $t_0$. The time-decayed weights can be written as
\[
w_{uv}(t)=e^{-\lambda(t-t_0)}\widetilde w_{uv},\qquad
\widetilde w_{uv}^{\mathrm{last}}=e^{\lambda(\tau_{uv}^{\mathrm{last}}-t_0)},\quad
\widetilde w_{uv}^{\mathrm{sum}}=\sum_{\tau\in\mathcal T_{uv}(t)}e^{\lambda(\tau-t_0)}.
\]
The common factor $e^{-\lambda(t-t_0)}$ cancels during row normalization. Algorithm~\ref{alg:incremental-ppr} summarizes the update for a selected configuration. It stores log-weights $a_{uv}$ and $b_{uv}$ for last and sum aggregation, respectively; an unobserved pair has both values $-\infty$. Setting $\lambda=0$ covers the no-decay case. Row normalization uses shifted log-weights for numerical stability.

\subsection{Computing Resources}
\label{app:computing-resources}

Our experiments were conducted on servers equipped with NVIDIA RTX A6000 and A100 GPUs. Both PPR variants use CPU-based historical graph maintenance and GPU-accelerated sparse matrix operations for finite-step propagation.

\FloatBarrier

\section{Additional Structural Analysis}
\label{app:structural-settings}

\begin{figure}[!htb]
\centering
\setlength{\abovecaptionskip}{4pt}
\setlength{\parskip}{0pt}
\includegraphics[width=\linewidth]{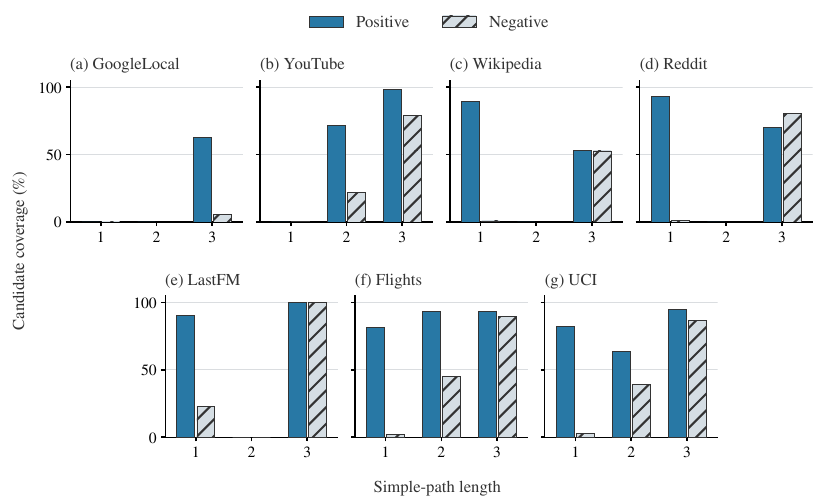}
\caption{Positive and negative candidate coverage by one-, two-, and three-step simple paths on seven complete test sets.}
\label{fig:structural-coverage-all}
\label{fig:structural-coverage}
\end{figure}

\textbf{Path Discriminability.} Figure~\ref{fig:structural-coverage-all} compares positive and negative candidate coverage by one-, two-, and three-step simple paths across seven datasets. For each path length, positive coverage is the fraction of test queries whose source is connected to the true destination by at least one such path. Negative coverage is the corresponding fraction of negative candidates connected to the query source.

Wikipedia, Reddit, LastFM, Flights, and UCI exhibit large positive--negative gaps in one-step coverage, indicating that direct historical interactions provide a strong recurrence signal. Longer paths are not uniformly more discriminative: three-step coverage is similar for positives and negatives on Wikipedia, higher for negatives on Reddit, and nearly complete for both on LastFM. In contrast, GoogleLocal has almost no direct historical links but a clear three-step coverage gap. YouTube shows a large two-step gap, while three-step paths reach many negative candidates as well. Flights and UCI also show two-step gaps. These patterns illustrate that useful path lengths depend on the dataset; broader reach alone does not ensure better candidate discrimination. Two-step coverage is zero between opposite partitions of a bipartite graph.

\begin{wraptable}{R}{0.52\linewidth}
\setlength{\abovecaptionskip}{0pt}
\input{tables/path_discriminability_scores}
\end{wraptable}
Table~\ref{tab:path-discriminability-scores} further evaluates candidate ranking by two- and three-step simple-path counts on the unweighted historical graphs available at prediction time. Repeated interactions are merged into a single edge, and no temporal weighting is applied. All reported values use complete test sets and the original candidates. Three-step path counts achieve AUCs of 80.14, 90.56, 95.77, and 94.60 on GoogleLocal, YouTube, Flickr, and WikiLink, respectively, showing that historical paths alone can distinguish positive from negative candidates. The useful path length varies across datasets: on Flights, two-step counts outperform three-step counts in both AUC and MRR.

\begin{figure}[p]
\centering
\includegraphics[width=\linewidth,height=0.9\textheight,keepaspectratio]{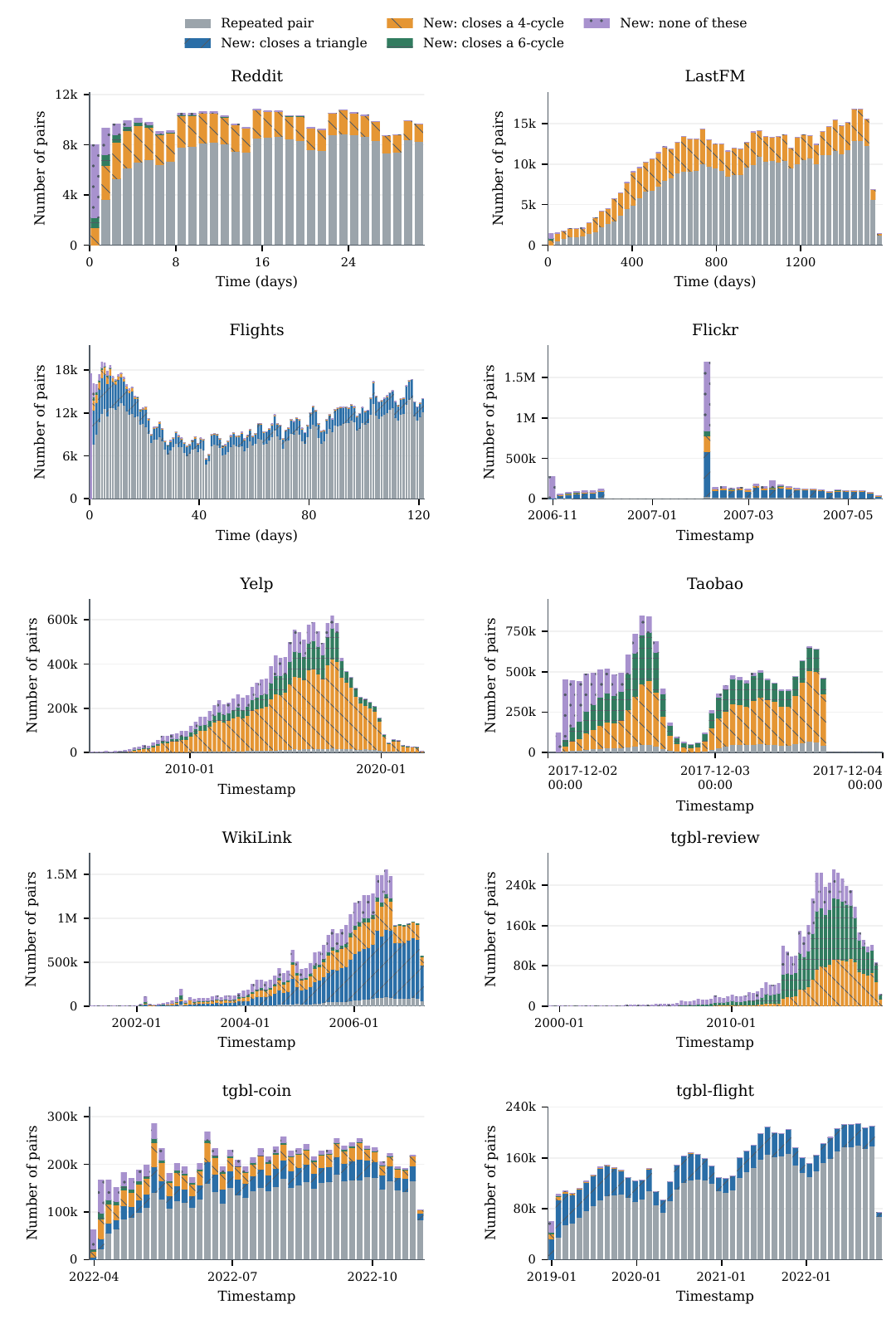}
\caption{Recurrence and historical cycle closure on the remaining ten datasets, using the categories and counting protocol of Figure~\ref{fig:event-structural-appearance}.}
\label{fig:event-structural-appearance-additional}
\end{figure}

\textbf{Structural Evolution Patterns.} Figures~\ref{fig:event-structural-appearance} and~\ref{fig:event-structural-appearance-additional} show how recurrence and cycle closure vary over time. Each undirected pair is counted once per display interval, at its first event. Using the unweighted full-history graph strictly before that timestamp, pairs are classified in order as repeated, new closing a triangle, new closing a four-cycle, new closing a six-cycle, or none of these. The three closure categories require simple paths of length two, three, and five, respectively. Equal-timestamp events are classified together before graph updates.

Repeated pairs dominate Reddit, LastFM, Flights, tgbl-coin, and tgbl-flight. Among new pairs, four-cycle closure is prominent on Reddit and LastFM, whereas triangle closure is more common on Flights and tgbl-flight; tgbl-coin shows both. Yelp and Taobao instead contain many new pairs closing four-cycles, together with a substantial six-cycle category. On tgbl-review, six-cycle closure is especially prominent, revealing connections through longer historical paths even when shorter closure conditions are absent. WikiLink contains many triangle-closing new pairs, alongside four-cycle closures. Flickr combines frequent triangle closure with a sharp activity burst in which many new pairs close none of the examined cycle types.

\section{Randomized Reference and Cycle Growth}
\label{app:cycle-growth}

This appendix expands the analysis of structural evolution in Section~\ref{sec:structural-discriminability} to help explain why PPR, as a multi-hop structural proximity measure, is effective for temporal link prediction. We compare observed cycle growth with a random ordering of the same edges to identify patterns in how historical connections develop. We present the random-order reference and its derivation, followed by the cycle-growth comparisons and final structural counts.

\begin{samepage}
\paragraph{Paths and cycle closure.}
A new interaction between previously unconnected nodes closes a $k$-cycle whenever its endpoints are connected by a $(k-1)$-step simple path. Specifically, for an absent edge $e=\{s,d\}$ in a simple undirected graph $G$ and $k\geq3$,
\[
N_k(G+e)-N_k(G)=S_{k-1}(s,d;G),
\]
where $N_k(G)$ counts simple $k$-cycles and $S_{k-1}(s,d;G)$ counts such paths. Each newly formed cycle corresponds to exactly one existing path between the endpoints. This relation connects cycle formation to the historical multi-hop paths examined in our prediction analysis. However, cycles can also accumulate under random edge arrival, motivating a reference against which to compare the observed growth.
\par\end{samepage}

\subsection{Random-Order Reference}
\label{app:random-order-reference}

To examine how edge arrival order shapes cycle formation, we compare the observed sequence with a reference that preserves the final graph while randomizing the order of its edges. Following the constrained-randomization framework of \citet{gauvin2022randomized}, let $G_M=(V,E_M)$ denote the final undirected simple graph constructed from the entire dataset, with repeated interactions merged and self-loops removed. Here, $M=|E_M|$ is the number of distinct edges. A uniformly random permutation of $E_M$ defines the reference sequence, and its first $m$ edges form $\widetilde G_m$. We compare the observed and randomized graphs at the same number of distinct edges, so repeated interactions do not advance this measure of graph growth.

Let $N_k(G)$ denote the number of simple $k$-cycles in $G$, counting each cycle once regardless of its starting vertex or traversal direction. A cycle in the final graph appears in $\widetilde G_m$ exactly when all of its $k$ edges occur among the first $m$ positions. For $m\geq k$, this probability is
\[
\Pr(c\subseteq\widetilde G_m\mid G_M)
=\frac{\binom{M-k}{m-k}}{\binom{M}{m}}
=\frac{\binom{m}{k}}{\binom{M}{k}}.
\]
Summing the indicator of this event over all final cycles gives
\[
\mathbb E[N_k(\widetilde G_m)\mid G_M]
=N_k(G_M)\frac{\binom{m}{k}}{\binom{M}{k}}.
\]
This follows from linearity of expectation and does not require independence between cycles. The expectation is zero when $m<k$.

This reference provides a baseline for when cycles emerge within a fixed final topology. Observed counts above the expectation indicate that the real edge ordering forms these cycles earlier in graph growth. The comparison is retrospective: the final graph defines the reference, while prediction uses the available historical graph.

\subsection{Cycle Growth Across Datasets}
\label{app:cycle-growth-results}

\begin{figure}[!htb]
\centering
\includegraphics[width=\linewidth]{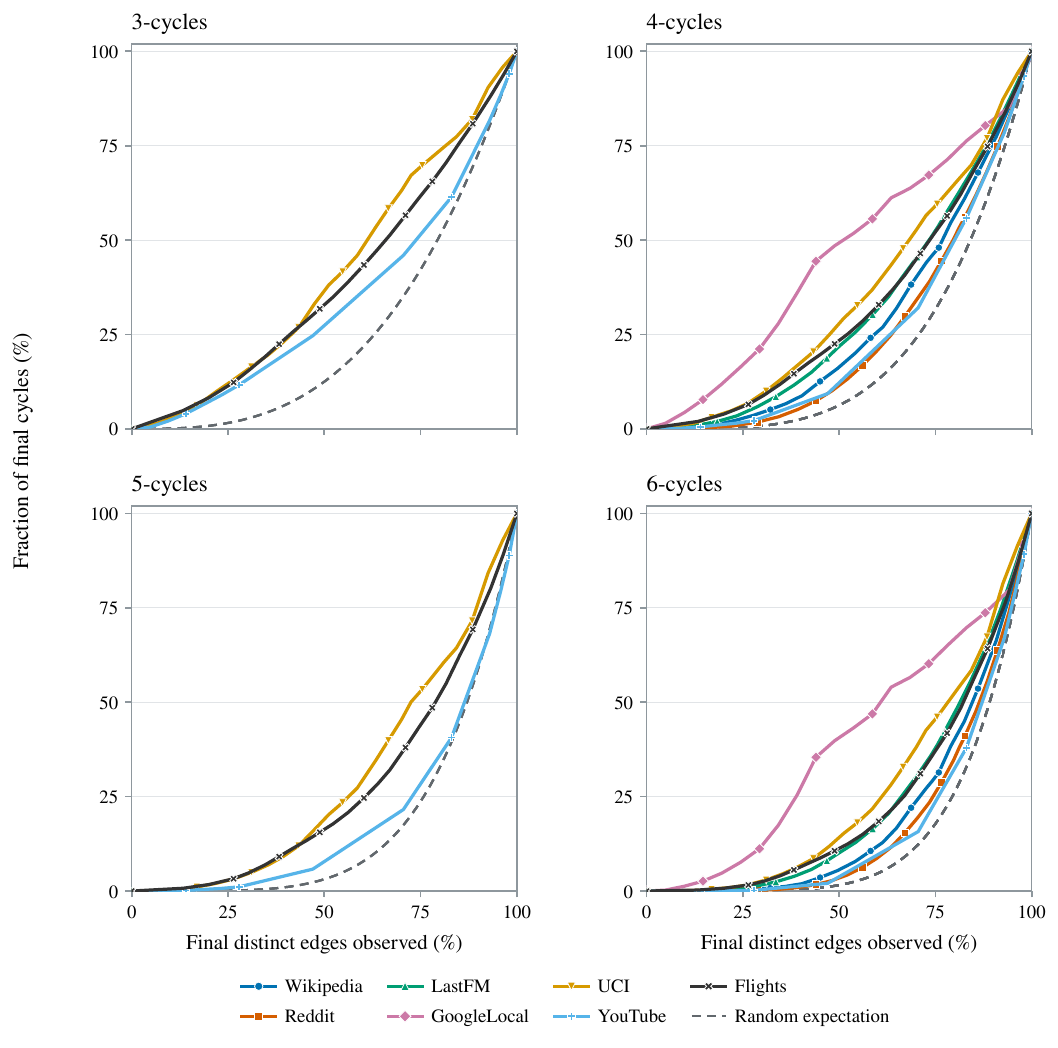}
\caption{Three- through six-cycle accumulation versus the exact random-order expectation, indexed by distinct-edge coverage.}
\label{fig:random-order-growth}
\end{figure}

\begin{figure}[!htb]
\centering
\includegraphics[width=\linewidth]{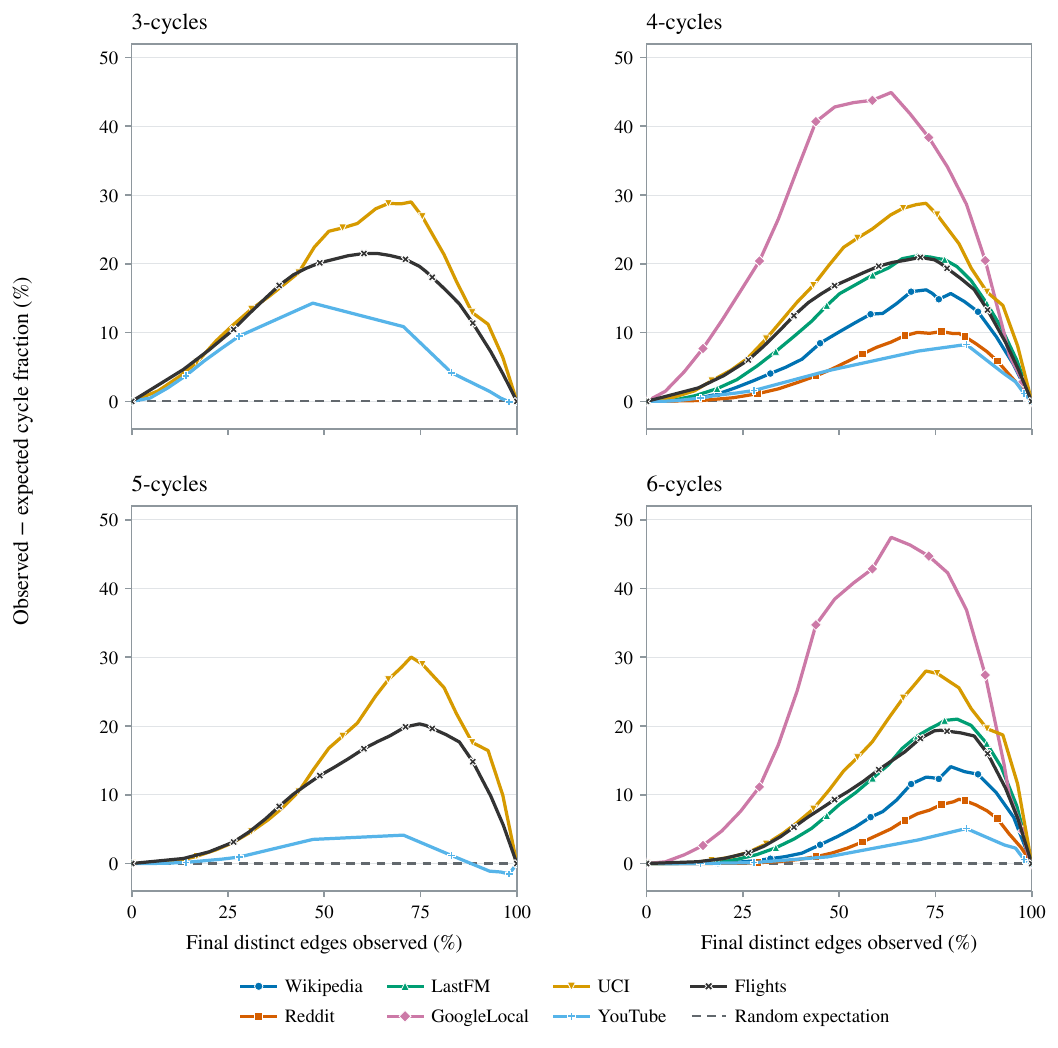}
\caption{Observed minus expected cycle fractions across datasets. Positive values indicate earlier cycle accumulation than under random edge ordering.}
\label{fig:random-order-excess}
\end{figure}

Using this reference, Figure~\ref{fig:random-order-growth} compares observed cycle accumulation with its expectation under random edge ordering, while Figure~\ref{fig:random-order-excess} shows the corresponding differences. Both figures track graph growth by the fraction of distinct edges observed, and normalize cycle counts by their totals in the final graph. Curves above the random-order expectation therefore indicate that a larger fraction of the final cycles has already formed at the same stage of graph growth. Across the examined datasets, four- and six-cycle counts exceed the expectation at all measured intermediate snapshots. Thus, the observed interaction order tends to bring together the edges forming these cycles earlier than a uniformly random ordering would.

The magnitude of this effect varies across datasets and cycle lengths. GoogleLocal exhibits particularly large deviations for both four- and six-cycles, indicating that these structures emerge well before the final graph is fully observed. UCI and Flights also show pronounced excesses across the examined cycle lengths. On YouTube, the deviation is more pronounced for triangles than for five-cycles, showing that different local structures need not develop in the same way. These observations connect graph evolution to the historical paths studied in our prediction analysis: when a new edge completes a cycle, its endpoints were already connected by an indirect path. The candidate-discriminability analysis in Section~\ref{sec:structural-discriminability} and Appendix~\ref{app:structural-settings} then examines whether such connections help distinguish future interaction targets from negative candidates.

\subsection{Final Structural Counts}

Table~\ref{tab:final-structural-counts} reports the total number of interactions and the structural counts of the final graph $G_M$, aggregated over the entire dataset. We merge repeated interactions into undirected edges and remove self-loops. A simple $k$-cycle contains $k$ distinct vertices and is counted once, regardless of its starting vertex or traversal direction. Counts depend on the aggregated topology, without requiring chronological order among interactions. The reported $M$ and $N_k(G_M)$ provide the final edge and cycle totals used in the preceding growth curves: intermediate counts are normalized by $N_k(G_M)$, and the random-order expectation is evaluated at the same distinct-edge progress $m/M$.

\input{tables/final_structural_counts}

\section{Structural Group Performance Decomposition}
\label{app:structural-group-decomposition}

We group test queries by how well three-step simple-path counts distinguish the true destination from negative candidates. Each candidate is scored by the number of such paths connecting it to the query source. In the strong-structure group, the true destination has strictly more paths than every negative candidate. Among the remaining queries with at least one path to the true destination, the medium-structure group contains those where it ranks within the top ten, and the weak-structure group contains those where it ranks below the top ten. Tied candidates receive their average rank. Queries with no three-step simple path to the true destination form the no-three-step-support group.

Using the main-table PPR configurations, we compare PPR and CRAFT on the same test queries and candidate sets. Table~\ref{tab:strong-structure-decomposition} reports each group's query share, MRR, and contribution to the overall performance gap. We calculate this contribution by multiplying the within-group MRR difference by the group's share of test queries; these contributions sum to the overall gap. Strong-structure queries account for nearly all of PPR's net gain on GoogleLocal and about half on YouTube.

\input{tables/strong_structure_decomposition}

\section{Sensitivity Settings and Additional Results}
\label{app:section63-settings}

\begin{figure}[!htb]
\centering
\includegraphics[width=\linewidth]{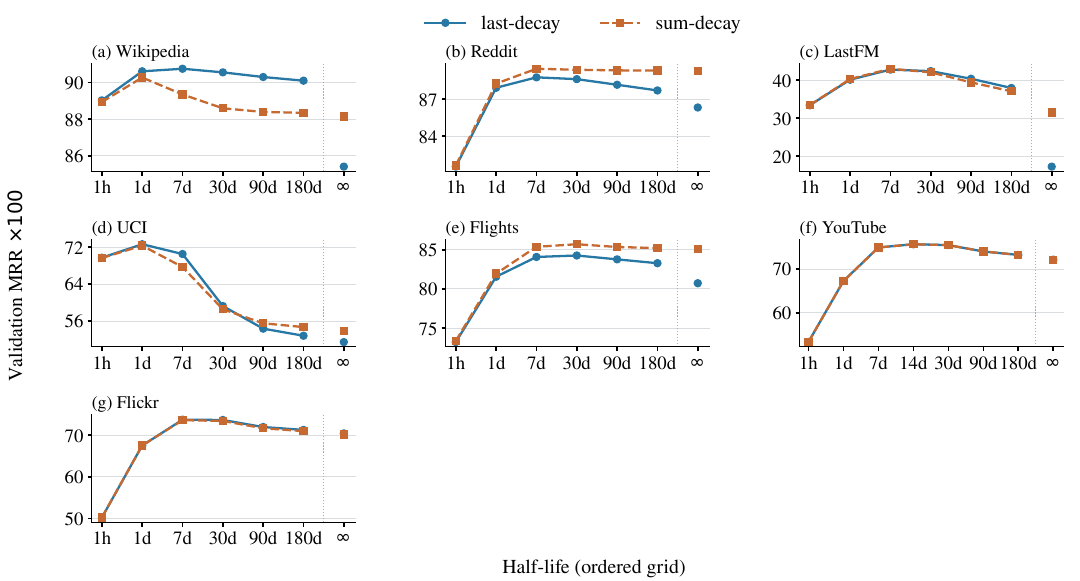}
\caption{Half-life and aggregation sensitivity on all seven datasets. Validation MRR $\times100$; panel-specific vertical scales.}
\label{fig:section63-temporal-all}
\end{figure}

\begin{figure}[!htbp]
\centering
\includegraphics[width=\linewidth]{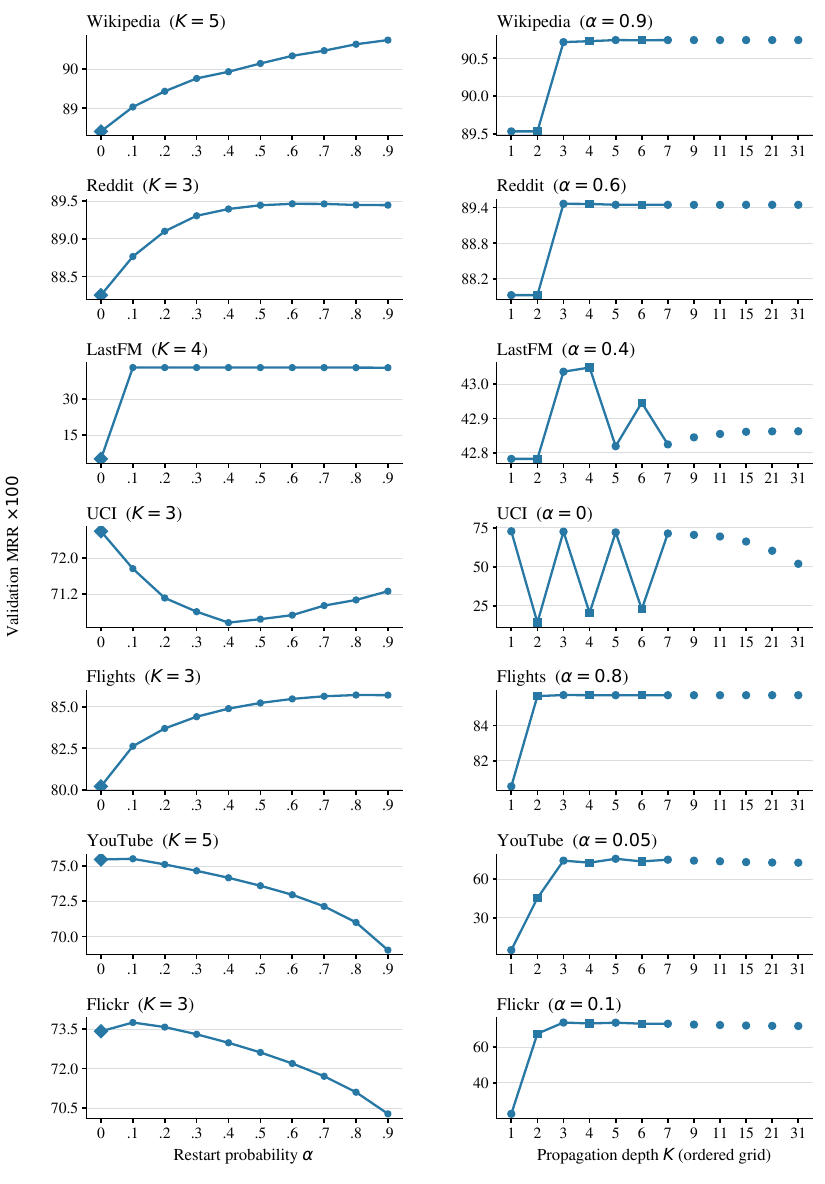}
\caption{Sensitivity to $\alpha$ and $K$ on seven datasets, with other parameters fixed. Validation MRR $\times100$.}
\label{fig:section63-propagation-all}
\end{figure}

Figures~\ref{fig:section63-temporal-all} and~\ref{fig:section63-propagation-all} report validation MRR across half-life, aggregation, restart probability, and propagation-depth sweeps. Each sweep varies the indicated parameters while keeping the others at the Time-decay PPR settings in Table~\ref{tab:selected-hyperparameters}. Half-life and $K$ use ordered grids, with the no-decay endpoint separated. Circles and squares denote odd and even $K$, and a diamond denotes $\alpha=0$.

For the window ablation in Table~\ref{tab:retention-four-arms}, we select the window length using validation MRR while retaining the aggregation and propagation settings of Time-decay PPR. Table~\ref{tab:retention-four-arm-settings} reports the selected windows and the mean fraction of historical events they exclude.

\input{tables/retention_four_arms_settings}

\paragraph{Temporal retention.}
Table~\ref{tab:retention-four-arm-settings} shows that the selected windows retain very different amounts of history across datasets. On Reddit, the selected window covers the entire available history, whereas on LastFM and UCI it excludes most historical events. A window length therefore does not correspond to a consistent information budget across datasets. Figure~\ref{fig:section63-temporal-all} further shows different responses to the decay time scale: UCI favors a short half-life and deteriorates as older interactions receive more weight, while Reddit and Flights maintain relatively stable performance over a broader range of longer half-lives. YouTube and Flickr perform poorly with a one-hour half-life but improve substantially when the time scale extends to days or weeks. These differences suggest that useful historical connections persist over different time scales across datasets.

\paragraph{Interaction between retention and aggregation.}
The effect of repeated-interaction aggregation also depends on temporal retention. On Wikipedia, last-decay outperforms sum-decay at every evaluated finite half-life, but this ordering reverses without decay. On LastFM, the two aggregation rules perform similarly at shorter half-lives, whereas sum aggregation performs substantially better in the full-history setting. These results show that the value of accumulating repeated interactions cannot be assessed independently of how their ages are weighted. By contrast, the nearly overlapping curves on YouTube and Flickr indicate that temporal weighting has a much larger effect than the choice between last and sum aggregation on these datasets.

\paragraph{Propagation depth and restart.}
Figure~\ref{fig:section63-propagation-all} shows that the benefit of additional propagation depends on the predictive information already available locally. Wikipedia and Reddit obtain further gains when moving from one to three steps, despite their strong one-step performance. YouTube and Flickr rely much more heavily on indirect connections, with substantial improvements over one-step propagation and their strongest results at shallow multi-step depths. Increasing depth further yields little benefit or reduces performance. Restart preferences also differ: larger restart probabilities perform well on Wikipedia and Flights, whereas YouTube and Flickr favor smaller values. Together, these patterns suggest that both the range of propagation and the balance between shorter and longer walks should reflect the dataset's interaction patterns.


\section{Additional Baseline and Candidate Analyses}
\label{app:additional-baselines-candidates}

\subsection{Additional Baselines}
\label{app:historical-baselines}

We compare Time-decay PPR with seven training-free baselines to examine how much predictive information can be obtained from direct interaction history and destination statistics.

\noindent\textbf{Pair-based baselines.} EdgeBank assigns a binary score indicating whether a source--destination pair has previously interacted \citep{poursafaei2022evaluation}. We evaluate full-history memory, EdgeBank$_\infty$, and windowed memory, EdgeBank$_{\mathrm{tw}}$. Following the paper's window setting, the latter retains interactions within a window whose length equals the validation time span. Pair-recency scores each candidate by its most recent interaction time with the query source, while pair-frequency uses their cumulative interaction count \citep{cornell2025heuristics}. These baselines preserve the source and destination roles of the original events.

\noindent\textbf{Destination-based baselines.} Destination-recency scores a candidate by its latest occurrence as a destination, irrespective of the query source. Destination-frequency counts all such occurrences \citep{cornell2025heuristics}. Time-aware popularity baselines have also been studied in recommendation and temporal graphs \citep{ji2020popularity,daniluk2023temporal}. Destination-activity instead weights them by recency:
\begin{equation}
a_h(v,t)=\sum_{(u,v,\tau)\in\mathcal E:\,\tau<t}2^{-(t-\tau)/h},
\end{equation}
where $h$ is the half-life. Thus, destination-frequency measures cumulative activity, whereas destination-activity emphasizes recent interactions. Candidates without relevant history receive zero for count-based scores and negative infinity for recency scores.

\noindent\textbf{Evaluation.} We evaluate all heuristics on the 14 main-table datasets using the original test candidate sets. Scores use only interactions strictly before the query timestamp, with history updated from observed positive events. The activity half-life is selected by validation MRR from $\{1\mathrm{h},1\mathrm{d},7\mathrm{d},14\mathrm{d},30\mathrm{d},90\mathrm{d},180\mathrm{d},\infty\}$. The other baselines require no parameter search. Time-decay PPR results use the configurations and refresh settings of the main tables.

\noindent\textbf{Results.} Table~\ref{tab:training-free-baselines} reports test MRR. Pair-recency and pair-frequency substantially improve upon binary edge memory on several seen-dominant datasets. On GoogleLocal, YouTube, and Flickr, destination activity provides stronger predictions than pair-based statistics, reaching MRRs of 20.01, 58.74, and 44.00, respectively, compared with 65.89, 70.56, and 71.93 for Time-decay PPR. On tgbl-review, destination activity reaches 41.58, exceeding Time-decay PPR and approaching the strongest reported neural baseline at 41.77. These comparisons illustrate the different predictive roles of relation recurrence and destination activity, alongside the broader competitiveness of historical structural proximity.

\begingroup
\makeatletter
\setlength{\@fptop}{0pt}
\setlength{\@fpsep}{14pt}
\setlength{\@fpbot}{0pt plus 1fil}
\makeatother
\input{tables/training_free_baselines}

\subsection{Activity-weighted Negative Sampling}
\label{app:activity-weighted-negative-sampling}

\noindent\textbf{Sampling strategy.} We retain the original test queries and positive destinations. For each query, we sample 100 distinct negative destinations without replacement, with each draw weighted by the remaining candidates' historical destination-activity scores defined in Appendix~\ref{app:historical-baselines}. We exclude all positive destinations associated with the same source and query timestamp. Sampling uses only historical interactions, and the resulting candidate sets are shared across methods and training seeds.

\noindent\textbf{Models and datasets.} We compare Time-decay PPR with JODIE, DyRep, TGN, GraphMixer, DyGFormer, and CRAFT. Table~\ref{tab:activity-weighted-results} reports test MRR under activity-weighted negative sampling on Wikipedia, Flights, GoogleLocal, YouTube, and Flickr.

\input{tables/activity_weighted_results}

\begin{samepage}
\noindent\textbf{Results and analysis.} Time-decay PPR remains competitive with the evaluated neural baselines under activity-weighted negative sampling. These results extend the comparison beyond the original benchmark candidates and show that historical structural proximity remains useful when negative candidates are drawn more frequently from recently active destinations. Together with the main results, this supports PPR as a competitive reference under different candidate distributions. The sensitivity of results to negative sampling further motivates evaluation protocols that assess model performance across multiple candidate distributions, providing a more comprehensive view of predictive capabilities.
\par\end{samepage}

\section{Implications for Future Research}
\label{app:future-research}

Our goal is not to suggest that structural proximity alone is sufficient for temporal link prediction or that PPR should replace learned models. Instead, PPR provides a simple, strong reference for understanding how far historical structural proximity can go and identifying where learned models offer additional benefits. Our findings motivate combining learned temporal representations with historical structural proximity, while adapting temporal retention and repeated-interaction aggregation to different interaction patterns. They also encourage evaluation under multiple negative sampling strategies to better characterize model strengths and limitations and guide future progress.

\clearpage
\endgroup

\end{document}

%% file: authors.tex
\author{%
Yuzhang Fei, Yanping Zheng, Lu Yi \& Zhewei Wei\thanks{Corresponding author: \texttt{zhewei@ruc.edu.cn}.}\\
Renmin University of China\\
\texttt{yuzhang.fei@ruc.edu.cn}, \texttt{zhengyanping@ruc.edu.cn}\\
\texttt{yilu@ruc.edu.cn}, \texttt{zhewei@ruc.edu.cn}
}

%% file: tables/main_results.tex

\begin{table}[!htb]
\centering
\caption{Test MRR $\times 100$ ($\uparrow$) on unseen-dominant datasets. Baseline results are from \citet{yi2025craft}.}
\label{tab:main-unseen}
\footnotesize
\setlength{\tabcolsep}{1pt}
\renewcommand{\arraystretch}{1.2}
\begin{tabular*}{\linewidth}{@{}l@{\extracolsep{\fill}}*{7}{c}@{}}
\toprule
Method & GoogleLocal & Yelp & Taobao & YouTube & Flickr & WikiLink & tgbl-review \\
\midrule
JODIE & \ResultCell{41.86}{\textpm 1.49} & \ResultCell{69.88}{\textpm 0.31} & \ResultCell{48.36}{\textpm 2.18} & \ResultCell{41.67}{\textpm 2.86} & \ResultCell{46.21}{\textpm 0.83} & \ResultCell{57.94}{\textpm 1.33} & \ResultCell{\RankSecond{41.43}}{\textpm 0.15} \\
DyRep & \ResultCell{37.73}{\textpm 1.34} & \ResultCell{57.69}{\textpm 1.05} & \ResultCell{40.03}{\textpm 2.40} & \ResultCell{35.12}{\textpm 4.13} & \ResultCell{38.04}{\textpm 4.19} & \ResultCell{42.63}{\textpm 1.33} & \ResultCell{\RankThird{40.06}}{\textpm 0.59} \\
TGAT & \ResultCell{19.78}{\textpm 0.24} & \ResultCell{---}{} & \ResultCell{---}{} & \ResultCell{43.56}{\textpm 2.53} & \ResultCell{23.53}{\textpm 3.35} & \ResultCell{---}{} & \ResultCell{19.64}{\textpm 0.23} \\
TGN & \ResultCell{54.13}{\textpm 1.97} & \ResultCell{69.79}{\textpm 0.24} & \ResultCell{60.28}{\textpm 0.54} & \ResultCell{55.16}{\textpm 5.89} & \ResultCell{46.03}{\textpm 6.78} & \ResultCell{62.94}{\textpm 2.16} & \ResultCell{37.48}{\textpm 0.23} \\
CAWN & \ResultCell{18.26}{\textpm 0.02} & \ResultCell{25.71}{\textpm 0.09} & \ResultCell{---}{} & \ResultCell{47.55}{\textpm 1.08} & \ResultCell{48.69}{\textpm 6.08} & \ResultCell{---}{} & \ResultCell{19.30}{\textpm 0.10} \\
GraphMixer & \ResultCell{21.31}{\textpm 0.14} & \ResultCell{33.96}{\textpm 0.19} & \ResultCell{31.54}{\textpm 0.02} & \ResultCell{58.87}{\textpm 0.12} & \ResultCell{45.01}{\textpm 0.08} & \ResultCell{48.57}{\textpm 0.02} & \ResultCell{36.89}{\textpm 1.50} \\
DyGFormer & \ResultCell{18.39}{\textpm 0.02} & \ResultCell{21.68}{\textpm 0.20} & \ResultCell{---}{} & \ResultCell{46.08}{\textpm 3.44} & \ResultCell{49.58}{\textpm 2.87} & \ResultCell{---}{} & \ResultCell{22.39}{\textpm 1.52} \\
CRAFT & \ResultCell{\RankThird{62.35}}{\textpm 0.46} & \ResultCell{\RankFirst{72.69}}{\textpm 0.61} & \ResultCell{\RankSecond{70.68}}{\textpm 0.42} & \ResultCell{\RankThird{58.92}}{\textpm 0.16} & \ResultCell{\RankThird{62.34}}{\textpm 0.84} & \ResultCell{\RankThird{75.48}}{\textpm 0.84} & \ResultCell{\RankFirst{41.77}}{\textpm 0.06} \\
\midrule
Window-based PPR & \ResultCell{\RankFirst{65.89}}{\textpm 0.00} & \ResultCell{\RankThird{71.31}}{\textpm 0.00} & \ResultCell{\RankThird{70.60}}{\textpm 0.00} & \ResultCell{\RankSecond{67.51}}{\textpm 0.00} & \ResultCell{\RankSecond{70.46}}{\textpm 0.00} & \ResultCell{\RankFirst{86.16}}{\textpm 0.00} & \ResultCell{28.20}{\textpm 0.00} \\
Time-decay PPR & \ResultCell{\RankFirst{65.89}}{\textpm 0.00} & \ResultCell{\RankSecond{72.28}}{\textpm 0.00} & \ResultCell{\RankFirst{71.81}}{\textpm 0.00} & \ResultCell{\RankFirst{70.56}}{\textpm 0.00} & \ResultCell{\RankFirst{71.93}}{\textpm 0.00} & \ResultCell{\RankFirst{86.16}}{\textpm 0.00} & \ResultCell{37.03}{\textpm 0.00} \\
\bottomrule
\end{tabular*}
\vspace{3pt}
\begin{minipage}{\linewidth}
\footnotesize
Mean $\pm$ standard deviation (deterministic PPR: $\pm0.00$). \RankFirst{1st}, \RankSecond{2nd}, and \RankThird{3rd} use displayed means; ties share ranks. A dash denotes an unreported result.
\end{minipage}
\end{table}

\begin{table}[!htb]
\centering
\caption{Test MRR $\times 100$ ($\uparrow$) on seen-dominant datasets. Baseline results are from \citet{yi2025craft}.}
\label{tab:main-seen}
\footnotesize
\setlength{\tabcolsep}{1pt}
\renewcommand{\arraystretch}{1.2}
\begin{tabular*}{\linewidth}{@{}l@{\extracolsep{\fill}}*{7}{c}@{}}
\toprule
Method & Wikipedia & Reddit & LastFM & UCI & Flights & tgbl-coin & tgbl-flight \\
\midrule
JODIE & \ResultCell{76.48}{\textpm 1.72} & \ResultCell{77.16}{\textpm 1.27} & \ResultCell{18.49}{\textpm 2.87} & \ResultCell{51.43}{\textpm 1.74} & \ResultCell{20.37}{\textpm 4.18} & \ResultCell{---}{} & \ResultCell{---}{} \\
DyRep & \ResultCell{67.42}{\textpm 4.21} & \ResultCell{72.33}{\textpm 2.14} & \ResultCell{17.76}{\textpm 6.56} & \ResultCell{14.00}{\textpm 2.71} & \ResultCell{15.62}{\textpm 0.47} & \ResultCell{45.20}{\textpm 4.60} & \ResultCell{55.60}{\textpm 1.40} \\
TGAT & \ResultCell{72.72}{\textpm 1.50} & \ResultCell{78.03}{\textpm 0.25} & \ResultCell{24.10}{\textpm 1.26} & \ResultCell{31.50}{\textpm 0.48} & \ResultCell{53.90}{\textpm 1.07} & \ResultCell{60.92}{\textpm 0.57} & \ResultCell{---}{} \\
TGN & \ResultCell{82.22}{\textpm 0.39} & \ResultCell{79.25}{\textpm 0.24} & \ResultCell{25.59}{\textpm 4.43} & \ResultCell{45.29}{\textpm 6.43} & \ResultCell{64.24}{\textpm 2.11} & \ResultCell{58.60}{\textpm 3.70} & \ResultCell{70.50}{\textpm 2.00} \\
CAWN & \ResultCell{86.07}{\textpm 0.08} & \ResultCell{87.66}{\textpm 0.04} & \ResultCell{35.10}{\textpm 0.34} & \ResultCell{66.96}{\textpm 0.48} & \ResultCell{80.06}{\textpm 2.10} & \ResultCell{---}{} & \ResultCell{---}{} \\
GraphMixer & \ResultCell{73.57}{\textpm 1.48} & \ResultCell{70.87}{\textpm 0.38} & \ResultCell{26.76}{\textpm 1.42} & \ResultCell{59.58}{\textpm 1.00} & \ResultCell{42.26}{\textpm 0.49} & \ResultCell{75.57}{\textpm 0.27} & \ResultCell{---}{} \\
DyGFormer & \ResultCell{\RankSecond{88.69}}{\textpm 0.04} & \ResultCell{88.70}{\textpm 0.05} & \ResultCell{\RankThird{46.50}}{\textpm 0.51} & \ResultCell{\RankSecond{76.61}}{\textpm 0.26} & \ResultCell{\RankThird{84.13}}{\textpm 3.46} & \ResultCell{75.17}{\textpm 0.38} & \ResultCell{---}{} \\
CRAFT-R & \ResultCell{\RankThird{88.25}}{\textpm 0.26} & \ResultCell{\RankThird{89.33}}{\textpm 0.11} & \ResultCell{\RankFirst{55.21}}{\textpm 0.23} & \ResultCell{\RankThird{75.11}}{\textpm 0.09} & \ResultCell{83.16}{\textpm 0.16} & \ResultCell{\RankFirst{88.47}}{\textpm 0.25} & \ResultCell{\RankFirst{91.39}}{\textpm 0.03} \\
\midrule
Window-based PPR & \ResultCell{86.01}{\textpm 0.00} & \ResultCell{\RankSecond{89.44}}{\textpm 0.00} & \ResultCell{44.85}{\textpm 0.00} & \ResultCell{71.54}{\textpm 0.00} & \ResultCell{\RankSecond{85.69}}{\textpm 0.00} & \ResultCell{\RankThird{79.03}}{\textpm 0.00} & \ResultCell{\RankThird{89.25}}{\textpm 0.00} \\
Time-decay PPR & \ResultCell{\RankFirst{89.17}}{\textpm 0.00} & \ResultCell{\RankFirst{89.61}}{\textpm 0.00} & \ResultCell{\RankSecond{46.66}}{\textpm 0.00} & \ResultCell{\RankFirst{78.80}}{\textpm 0.00} & \ResultCell{\RankFirst{86.15}}{\textpm 0.00} & \ResultCell{\RankSecond{81.29}}{\textpm 0.00} & \ResultCell{\RankSecond{89.45}}{\textpm 0.00} \\
\bottomrule
\end{tabular*}
\vspace{3pt}
\begin{minipage}{\linewidth}
\footnotesize
Mean $\pm$ standard deviation (deterministic PPR: $\pm0.00$). \RankFirst{1st}, \RankSecond{2nd}, and \RankThird{3rd} use displayed means; ties share ranks. A dash denotes an unreported result.
\end{minipage}
\end{table}

%% file: tables/retention_four_arms.tex
\centering
\caption{Ablations of Time-decay PPR (validation MRR $\times100$).}
\label{tab:retention-four-arms}
\small
\setlength{\tabcolsep}{3pt}
\begin{tabular*}{\linewidth}{@{}l@{\extracolsep{\fill}}rrr@{}}
\toprule
Dataset & \shortstack{Time-decay\\PPR} & \shortstack{w. window,\\w/o decay} & \shortstack{w/o\\decay} \\
\midrule
Wikipedia & \textbf{90.74} & 86.69 & 85.42 \\
Reddit$\dagger$ & \textbf{89.46} & 89.28 & 89.28 \\
LastFM & \textbf{43.05} & 40.68 & 31.46 \\
UCI & \textbf{72.60} & 66.26 & 51.42 \\
Flights & \textbf{85.71} & 85.16 & 85.06 \\
YouTube & \textbf{75.64} & 71.21 & 72.11 \\
Flickr & \textbf{73.75} & 70.78 & 70.46 \\
tgbl-coin & \textbf{78.69} & 75.26 & 71.08 \\
\bottomrule
\end{tabular*}
\par\smallskip
{\footnotesize $\dagger$ The selected window covers all history. Paired settings are given in Appendix~\ref{app:section63-settings}.}

%% file: tables/dataset_statistics.tex
\begin{table}[htbp]
\centering
\caption{Dataset statistics reported by \citet{yi2025craft}.}
\label{tab:dataset-statistics}
\footnotesize
\setlength{\tabcolsep}{3pt}
\renewcommand{\arraystretch}{1.13}
\begin{tabular*}{\linewidth}{@{}l@{\extracolsep{\fill}}lrrrr@{}}
\toprule
Dataset & Interaction & Nodes & Events & Timestamps & Repeat (\%) \\
\midrule
\multicolumn{6}{@{}l}{\textit{Unseen-dominant}} \\
GoogleLocal & Business review & 206,244/267,336 & 1,913,967 & 1,771,060 & 0.00 \\
Yelp & Business review & 1,338,688/405,081 & 19,760,293 & 14,646,734 & 25.18 \\
Taobao & Product interaction & 760,617/863,016 & 18,853,792 & 139,171 & 16.58 \\
YouTube & Follow & 402,422 & 3,288,028 & 203 & 0.00 \\
Flickr & Follow & 233,836 & 7,223,559 & 134 & 0.00 \\
WikiLink & Hyperlink & 1,361,972 & 34,163,774 & 2,198 & 0.00 \\
tgbl-review & Product review & 352,636/298,590 & 4,873,540 & 6,865 & 0.19 \\
\midrule
\multicolumn{6}{@{}l}{\textit{Seen-dominant}} \\
Wikipedia & Page edit & 8,227/1,000 & 157,474 & 152,757 & 88.41 \\
Reddit & Subreddit interaction & 10,000/984 & 672,447 & 669,065 & 88.32 \\
LastFM & Song listening & 980/1,000 & 1,293,103 & 1,283,614 & 88.01 \\
UCI & Message & 1,899 & 59,835 & 58,911 & 66.06 \\
Flights & Flight & 13,169 & 1,927,145 & 122 & 79.50 \\
tgbl-coin & Transaction & 638,486 & 22,809,486 & 1,295,720 & 82.93 \\
tgbl-flight & Flight & 18,143 & 67,169,570 & 1,385 & 96.48 \\
\bottomrule
\end{tabular*}
\vspace{3pt}
\begin{minipage}{\linewidth}
\footnotesize
Bipartite node counts are source/destination counts; timestamps are distinct. Repeat is the percentage of events whose ordered pair appeared strictly earlier \citep{yi2025tgbseq}.
\end{minipage}
\end{table}

%% file: tables/selected_hyperparameters.tex
\begin{table}[!htb]
\centering
\caption{Configurations used for the main-table PPR results.}
\label{tab:selected-hyperparameters}
\small
\setlength{\tabcolsep}{4pt}
\renewcommand{\arraystretch}{1.12}
\begin{tabular}{@{}lcccccccc@{}}
\toprule
& \multicolumn{4}{c}{Window-based PPR} & \multicolumn{4}{c}{Time-decay PPR} \\
\cmidrule(lr){2-5}\cmidrule(l){6-9}
Dataset & Aggr. & $W$ & $\alpha$ & $K$ & Aggr. & $h$ & $\alpha$ & $K$ \\
\midrule
Wikipedia & count & $7\,\mathrm{d}$ & 0.6 & 7 & last & $7\,\mathrm{d}$ & 0.9 & 5 \\
Reddit & count & $\infty$ & 0.6 & 3 & sum & $7\,\mathrm{d}$ & 0.6 & 3 \\
LastFM & count & $30\,\mathrm{d}$ & 0.8 & 4 & sum & $7\,\mathrm{d}$ & 0.4 & 4 \\
UCI & count & $7\,\mathrm{d}$ & 0 & 3 & last & $1\,\mathrm{d}$ & 0 & 3 \\
Flights & count & $90\,\mathrm{d}$ & 0.9 & 3 & sum & $30\,\mathrm{d}$ & 0.8 & 3 \\
GoogleLocal & binary & $\infty$ & 0.1 & 9 & sum & $\infty$ & 0.1 & 9 \\
YouTube & count & $\infty$ & 0.05 & 3 & last & $14\,\mathrm{d}$ & 0.05 & 5 \\
Flickr & binary & $90\,\mathrm{d}$ & 0.1 & 3 & last & $7\,\mathrm{d}$ & 0.1 & 3 \\
Yelp & binary & $\infty$ & 0.1 & 5 & last & $180\,\mathrm{d}$ & 0.05 & 7 \\
Taobao & binary & $\infty$ & 0.1 & 5 & last & $1\,\mathrm{d}$ & 0.1 & 5 \\
WikiLink & count & $\infty$ & 0.2 & 5 & sum & $\infty$ & 0.2 & 5 \\
tgbl-review & binary & $\infty$ & 0 & 31 & sum & $90\,\mathrm{d}$ & 0 & 31 \\
tgbl-coin & count & $30\,\mathrm{d}$ & 0.7 & 7 & sum & $1\,\mathrm{d}$ & 0.9 & 9 \\
tgbl-flight & count & $30\,\mathrm{d}$ & 0.9 & 3 & sum & $30\,\mathrm{d}$ & 0.2 & 1 \\
\bottomrule
\end{tabular}
\end{table}

%% file: tables/path_discriminability_scores.tex
\centering
\small
\caption{Query-averaged ROC-AUC and MRR ($\times 100$) of two- and three-step simple-path counts on the test set.}
\label{tab:path-discriminability-scores}
\setlength{\tabcolsep}{7pt}
\begin{tabular}{@{}lrrrr@{}}
\toprule
& \multicolumn{2}{c}{2-hop} & \multicolumn{2}{c}{3-hop} \\
\cmidrule(lr){2-3}\cmidrule(lr){4-5}
Dataset & AUC & MRR & AUC & MRR \\
\midrule
\multicolumn{5}{@{}l}{\textit{Bipartite graphs}} \\[2pt]
Wikipedia & --- & --- & 58.75 & 18.05 \\
Reddit & --- & --- & 70.59 & 28.04 \\
LastFM & --- & --- & 53.54 & 8.66 \\
GoogleLocal & --- & --- & 80.14 & 48.61 \\
Yelp & --- & --- & 92.38 & 63.16 \\
Taobao & --- & --- & 86.92 & 59.54 \\
tgbl-review & --- & --- & 70.47 & 20.93 \\
\midrule
\multicolumn{5}{@{}l}{\textit{Homogeneous graphs}} \\[2pt]
UCI & 66.54 & 12.58 & 81.15 & 21.39 \\
Flights & 93.71 & 67.50 & 90.69 & 56.83 \\
YouTube & 79.04 & 36.21 & 90.56 & 58.16 \\
Flickr & 91.83 & 60.41 & 95.77 & 58.18 \\
WikiLink & 88.38 & 62.45 & 94.60 & 66.91 \\
tgbl-flight & 79.99 & 59.73 & 76.05 & 51.50 \\
\bottomrule
\end{tabular}
\par\smallskip
\begin{minipage}{\linewidth}\footnotesize
---: omitted because source and destination nodes in opposite partitions of a bipartite graph have no two-step paths. AUC gives half credit to ties; MRR uses average ranks for tied scores.
\end{minipage}

%% file: tables/final_structural_counts.tex
\begin{table}[!htbp]
\centering
\caption{Interactions, distinct edges, and simple cycles in the final aggregated graph.}
\label{tab:final-structural-counts}
\small
\setlength{\tabcolsep}{4pt}
\begin{tabular}{@{}lrrrrrr@{}}
\toprule
Dataset & \multicolumn{1}{c}{Interactions} & \multicolumn{1}{c}{Distinct edges $M$} & \multicolumn{1}{c}{3-cycles} & \multicolumn{1}{c}{4-cycles} & \multicolumn{1}{c}{5-cycles} & \multicolumn{1}{c}{6-cycles} \\
\midrule
GoogleLocal & 1,913,967 & 1,913,967 & $0\phantom{.00\times10^{00}}$ & $2.03\times10^{7\phantom{0}}$ & $0\phantom{.00\times10^{00}}$ & $3.26\times10^{10}$ \\
YouTube & 3,288,028 & 3,288,028 & $6.57\times10^{6\phantom{0}}$ & $5.96\times10^{9\phantom{0}}$ & $3.90\times10^{11}$ & $3.83\times10^{14}$ \\
Wikipedia & 157,474 & 18,257 & $0\phantom{.00\times10^{00}}$ & $1.22\times10^{5\phantom{0}}$ & $0\phantom{.00\times10^{00}}$ & $3.38\times10^{7\phantom{0}}$ \\
Reddit & 672,447 & 78,516 & $0\phantom{.00\times10^{00}}$ & $1.25\times10^{7\phantom{0}}$ & $0\phantom{.00\times10^{00}}$ & $4.15\times10^{10}$ \\
LastFM & 1,293,103 & 154,993 & $0\phantom{.00\times10^{00}}$ & $5.98\times10^{8\phantom{0}}$ & $0\phantom{.00\times10^{00}}$ & $1.85\times10^{13}$ \\
Flights & 1,927,145 & 295,795 & $3.94\times10^{6\phantom{0}}$ & $5.08\times10^{8\phantom{0}}$ & $8.15\times10^{10}$ & $1.47\times10^{13}$ \\
UCI & 59,835 & 13,838 & $1.43\times10^{4\phantom{0}}$ & $7.29\times10^{5\phantom{0}}$ & $1.85\times10^{7\phantom{0}}$ & $8.76\times10^{8\phantom{0}}$ \\
\bottomrule
\end{tabular}
\end{table}

%% file: tables/strong_structure_decomposition.tex
\begin{table}[!htbp]
\centering
\caption{Test MRR $\times100$ by structural group. $\Delta$ is PPR minus CRAFT; contribution is query share times $\Delta$.}
\label{tab:strong-structure-decomposition}
\small
\setlength{\tabcolsep}{4pt}
\begin{tabular}{llrrrrr}
\toprule
Dataset & Group & Share (\%) & PPR & CRAFT & $\Delta$ & Contribution \\
\midrule
GoogleLocal & Strong & 38.85 & 94.63 & 86.61 & +8.02 & +3.12 \\
GoogleLocal & Medium & 22.37 & 59.09 & 57.43 & +1.65 & +0.37 \\
GoogleLocal & Weak & 1.34 & 23.11 & 32.42 & -9.31 & -0.12 \\
GoogleLocal & No 3-step & 37.44 & 41.65 & 42.24 & -0.60 & -0.22 \\
GoogleLocal & All & 100.00 & 65.89 & 62.75 & +3.14 & +3.14 \\
\midrule
YouTube & Strong & 49.13 & 96.99 & 85.48 & +11.51 & +5.66 \\
YouTube & Medium & 25.83 & 62.18 & 44.83 & +17.35 & +4.48 \\
YouTube & Weak & 23.49 & 27.47 & 21.71 & +5.76 & +1.35 \\
YouTube & No 3-step & 1.55 & 25.29 & 25.63 & -0.34 & -0.01 \\
YouTube & All & 100.00 & 70.56 & 59.07 & +11.49 & +11.49 \\
\bottomrule
\end{tabular}
\end{table}

%% file: tables/retention_four_arms_settings.tex
\begin{table}[!htb]
\centering
\caption{Selected windows and mean fractions of excluded historical events.}
\label{tab:retention-four-arm-settings}
\small
\begin{tabular}{lrr}
\toprule
Dataset & Window & Excluded events \\
\midrule
Wikipedia & 7d & 68.47\% \\
Reddit & 90d & 0.00\% \\
LastFM & 30d & 96.93\% \\
UCI & 7d & 90.92\% \\
Flights & 90d & 13.53\% \\
YouTube & 90d & 31.95\% \\
Flickr & 90d & 18.92\% \\
tgbl-coin & 30d & 79.49\% \\
\bottomrule
\end{tabular}
\end{table}

%% file: tables/training_free_baselines.tex
\begin{table}[!htbp]
\centering
\caption{Test MRR $\times100$ of heuristic baselines on the original candidate sets. Time-decay PPR results follow the main tables.}
\label{tab:training-free-baselines}
\begingroup
\footnotesize
\setlength{\tabcolsep}{1pt}
\renewcommand{\arraystretch}{1.12}
\begin{tabular*}{\linewidth}{@{}l@{\extracolsep{\fill}}*{7}{c}@{}}
\toprule
 & \multicolumn{7}{c}{\textit{Unseen-dominant}} \\
\cmidrule(lr){2-8}
Method & GoogleLocal & Yelp & Taobao & YouTube & Flickr & WikiLink & tgbl-review \\
\midrule
EdgeBank$_\infty$ & 1.96 & 9.77 & 20.28 & 1.95 & 1.96 & 1.96 & 1.89 \\
EdgeBank$_{\mathrm{tw}}$ & 1.96 & 5.89 & 14.62 & 1.96 & 1.96 & 1.96 & 2.05 \\
Pair-recency & 1.96 & 9.79 & 20.29 & 1.95 & 1.96 & 1.96 & 1.97 \\
Pair-frequency & 1.96 & 9.78 & 20.29 & 1.95 & 1.96 & 1.96 & 1.89 \\
\midrule
Destination-recency & 13.41 & 24.78 & 20.72 & 34.57 & 20.78 & 30.55 & \RankThird{33.01} \\
Destination-frequency & \RankThird{18.39} & \RankThird{24.79} & \RankThird{30.03} & \RankThird{51.31} & \RankThird{32.37} & \RankThird{47.50} & 28.06 \\
Destination-activity & \RankSecond{20.01} & \RankSecond{34.06} & \RankSecond{30.22} & \RankSecond{58.74} & \RankSecond{44.00} & \RankSecond{48.54} & \RankFirst{41.58} \\
\midrule
Time-decay PPR & \RankFirst{65.89} & \RankFirst{72.28} & \RankFirst{71.81} & \RankFirst{70.56} & \RankFirst{71.93} & \RankFirst{86.16} & \RankSecond{37.03} \\
\bottomrule
\end{tabular*}
\par\vspace{7pt}
\begin{tabular*}{\linewidth}{@{}l@{\extracolsep{\fill}}*{7}{c}@{}}
\toprule
 & \multicolumn{7}{c}{\textit{Seen-dominant}} \\
\cmidrule(lr){2-8}
Method & Wikipedia & Reddit & LastFM & UCI & Flights & tgbl-coin & tgbl-flight \\
\midrule
EdgeBank$_\infty$ & 79.56 & 77.89 & 14.79 & 42.57 & 54.46 & 35.99 & 16.69 \\
EdgeBank$_{\mathrm{tw}}$ & 81.03 & 78.46 & 18.59 & 53.77 & 58.98 & 57.36 & 37.57 \\
Pair-recency & \RankSecond{87.49} & \RankThird{87.22} & \RankSecond{43.62} & \RankSecond{66.87} & \RankThird{75.91} & \RankSecond{78.23} & \RankThird{83.24} \\
Pair-frequency & \RankThird{84.84} & \RankSecond{87.79} & \RankThird{36.72} & \RankThird{56.51} & \RankSecond{77.97} & 70.07 & \RankSecond{87.23} \\
\midrule
Destination-recency & 48.97 & 35.51 & 18.92 & 50.26 & 5.19 & 72.63 & 18.31 \\
Destination-frequency & 12.22 & 47.28 & 7.70 & 17.92 & 40.59 & 61.30 & 60.27 \\
Destination-activity & 42.15 & 48.10 & 21.91 & 50.60 & 41.57 & \RankThird{72.95} & 63.46 \\
\midrule
Time-decay PPR & \RankFirst{89.17} & \RankFirst{89.61} & \RankFirst{46.66} & \RankFirst{78.80} & \RankFirst{86.15} & \RankFirst{81.29} & \RankFirst{89.45} \\
\bottomrule
\end{tabular*}
\par\vspace{3pt}
\begin{minipage}{\linewidth}
\footnotesize
EdgeBank$_{\mathrm{tw}}$ uses a fixed validation-duration window; activity uses validation-selected decay. \RankFirst{First}, \RankSecond{second}, and \RankThird{third} are marked per dataset.
PPR results are those in the main comparison.
\end{minipage}
\endgroup
\end{table}

%% file: tables/activity_weighted_results.tex
\begin{table}[!htb]
\centering
\caption{Test MRR $\times100$ ($\uparrow$) under activity-weighted negative sampling (mean $\pm$ standard deviation over three seeds).}
\label{tab:activity-weighted-results}
\footnotesize
\setlength{\tabcolsep}{1pt}
\renewcommand{\arraystretch}{1.2}
\begin{tabular*}{\linewidth}{@{}l@{\extracolsep{\fill}}*{5}{c}@{}}
\toprule
Method & Wikipedia & Flights & GoogleLocal & YouTube & Flickr \\
\midrule
JODIE & \ResultCell{66.39}{\textpm 0.51} & \ResultCell{11.06}{\textpm 2.00} & \ResultCell{25.63}{\textpm 0.12} & \ResultCell{5.88}{\textpm 0.78} & \ResultCell{17.60}{\textpm 1.47} \\
DyRep & \ResultCell{62.39}{\textpm 1.66} & \ResultCell{5.54}{\textpm 0.24} & \ResultCell{22.29}{\textpm 0.28} & \ResultCell{5.03}{\textpm 0.21} & \ResultCell{14.34}{\textpm 0.34} \\
TGN & \ResultCell{77.47}{\textpm 0.92} & \ResultCell{30.57}{\textpm 1.46} & \ResultCell{\RankThird{37.35}}{\textpm 0.18} & \ResultCell{\RankSecond{11.04}}{\textpm 1.34} & \ResultCell{19.20}{\textpm 1.03} \\
GraphMixer & \ResultCell{60.93}{\textpm 0.17} & \ResultCell{11.95}{\textpm 0.28} & \ResultCell{7.01}{\textpm 0.05} & \ResultCell{4.84}{\textpm 0.15} & \ResultCell{4.82}{\textpm 0.18} \\
DyGFormer & \ResultCell{\RankSecond{85.16}}{\textpm 0.08} & \ResultCell{\RankSecond{60.89}}{\textpm 0.21} & \ResultCell{4.77}{\textpm 0.20} & \ResultCell{5.73}{\textpm 0.50} & \ResultCell{\RankThird{22.46}}{\textpm 3.11} \\
CRAFT / CRAFT-R & \ResultCell{\RankThird{83.40}}{\textpm 0.12} & \ResultCell{\RankThird{56.65}}{\textpm 0.58} & \ResultCell{\RankFirst{49.54}}{\textpm 0.50} & \ResultCell{\RankThird{10.83}}{\textpm 0.56} & \ResultCell{\RankSecond{29.12}}{\textpm 0.32} \\
\midrule
Time-decay PPR & \ResultCell{\RankFirst{85.38}}{\textpm 0.00} & \ResultCell{\RankFirst{60.92}}{\textpm 0.00} & \ResultCell{\RankSecond{47.96}}{\textpm 0.00} & \ResultCell{\RankFirst{14.04}}{\textpm 0.00} & \ResultCell{\RankFirst{33.43}}{\textpm 0.00} \\
\bottomrule
\end{tabular*}
\vspace{3pt}
\begin{minipage}{\linewidth}\footnotesize
CRAFT-R is used on Wikipedia and Flights; CRAFT is used on GoogleLocal, YouTube, and Flickr.
\end{minipage}
\end{table}